\documentclass[traditabstract]{aa} % for a referee version
\usepackage{float,lscape}
\usepackage{rotating}
\usepackage{varwidth}
\usepackage{textpos}
\usepackage{rotating} % Rotating table
\usepackage{graphicx}
\usepackage{natbib}
\usepackage{amsmath}
\usepackage{mwe,tikz}
\usepackage[percent]{overpic}
\renewcommand{\arraystretch}{1.3}
\usepackage[para,online,flushleft]{threeparttable}
\usepackage{txfonts}
\usepackage{hyperref}
\def\eflux{{\rm erg~cm$^{-2}$~s$^{-1}$}}

\def\src{IGR\,J17544-2619}
\def\inte{{\em INTEGRAL}}
\def\xmm{{\em XMM-Newton}}

\def\swift{{\em Swift}}

\def \inte {{\em INTEGRAL}}
\def \xmm {{\em XMM--Newton}}
\def \nustar {{\em NuSTAR}}

\def \ergsec{\hbox{erg s$^{-1}$}}

\def \hcm {\hbox {\ifmmode $ atom cm$^{-2}\else atom cm$^{-2}$\fi}}

\def \apjl {ApJL}
\def \apjs {ApJS}\def \aap {A\&A}

\begin{document}

   \title{Multi-wavelength observations of IGR\,J17544-2619 from quiescence to outburst}

   \author{E. Bozzo
    \inst{1}
    \and V. Bhalerao
    \inst{2}
   \and P. Pradhan
     \inst{3,4}        
   \and J. Tomsick
     \inst{5}      
   \and P. Romano
     \inst{6} 
    \and C. Ferrigno 
    \inst{1}
    \and S. Chaty
     \inst{7,8} 
     \and L. Oskinova
     \inst{9} 
     \and A. Manousakis
      \inst{10}   
     \and R. Walter 
     \inst{1} 
     \and M. Falanga 
      \inst{11,12}   
     \and S. Campana 
      \inst{13}   
     \and L. Stella 
      \inst{14}   
      \and M. Ramolla 
      \inst{15}   
      \and R. Chini
      \inst{15,16}   
    }

   \institute{ISDC Data Centre for Astrophysics, Chemin d’Ecogia 16, CH-1290 Versoix, Switzerland; \email{enrico.bozzo@unige.ch}
    \and   
    Inter-University Center for Astronomy and Astrophysics, Post Bag 4, Ganeshkhind, Pune 411007, India
    \and  
    St. Joseph’s College, Singamari, Darjeeling-734104, West Bengal, India
    \and    
    North Bengal University, Raja Rammohanpur, District Darjeeling-734013, West Bengal, India
    \and 
    Space Sciences Laboratory, 7 Gauss Way, University of California, Berkeley, CA 94720-7450, USA
    \and    
    INAF, Istituto di Astrofisica Spaziale e Fisica Cosmica - Palermo, via U. La Malfa 153, 90146 Palermo, Italy 
    \and 
    Laboratoire AIM (UMR 7158, CEA/DRF/Irfu/SAp-CNRS-Universit\'e Paris Diderot), Centre de Saclay, L’Orme des Merisiers, B\^{a}t. 709, 
    FR-91191 Gif-sur-Yvette Cedex, France
    \and
    Institut Universitaire de France, 103, boulevard Saint-Michel, 75005 Paris, France  
    \and 
    Institut f\"ur Physik und Astronomie, Universit\"at Potsdam, Karl-Liebknecht-Str. 24/25, D-14476 Potsdam, Germany  
    \and 
    Centrum Astronomiczne im.\ M. Kopernika, Bartycka 18, 00-716 Warszawa, Poland 
    \and 
    International Space Science Institute (ISSI), Hallerstrasse 6, CH-3012 Bern, Switzerland
    \and 
    International Space Science Institute in Beijing, No. 1 Nan Er Tiao, Zhong Guan Cun, Beijing 100190, China
    \and 
    INAF - Osservatorio Astronomico di Brera, via Emilio Bianchi 46, I-23807 Merate (LC), Italy. 
    \and     
    INAF - Osservatorio Astronomico di Roma, Via Frascati 33, 00044 Rome, Italy. 
    \and     
    Ruhr-Universit\"at Bochum, 44780 Bochum, Germany
    \and 
    Instituto de Astronomía, Universidad Cat\'olica del Norte, Avenida Angamos 0610,  Antofagasta, Chile
}

   \date{}

  \abstract{In this paper we report on a long multi-wavelength observational campaign of the supergiant fast X-ray transient 
  prototype \src.\ A 150~ks-long observation was carried out simultaneously with \xmm\ and \nustar,\ catching the 
  source in an initial faint X-ray state and then undergoing a bright X-ray outburst lasting approximately 7~ks. We studied the spectral variability 
  during outburst and quiescence by using a thermal and bulk Comptonization model that is typically adopted to describe the X-ray spectral 
  energy distribution of young pulsars in high mass X-ray binaries. Although the statistics of the collected X-ray data were relatively high, 
  we could neither confirm the presence of a cyclotron line in the broad-band spectrum of the source (0.5-40~keV), nor 
  detect any of the previously reported tentative detections of the source spin period.  
  The monitoring carried out with \swift\,/XRT during the same orbit of the system observed by \xmm\ and \nustar\ revealed 
  that the source remained in a low emission state for most of the time, in agreement with the known property of  
  all supergiant fast X-ray transients being significantly sub-luminous compared to other supergiant X-ray binaries. 
  Optical and infrared observations were carried out for a total of a few thousand seconds during the quiescence state of the source detected by \xmm\ and \nustar.\ The measured optical and infrared 
  magnitudes were slightly lower than previous values reported in the literature, but compatible with the known micro-variability 
  of supergiant stars. UV observations obtained with the UVOT telescope on-board \swift\ did not reveal significant changes 
  in the magnitude of the source in this energy domain compared to previously reported values.}   
  

   \maketitle

\section{Introduction}
\label{sec:intro}

Supergiant Fast X-ray Transients (SFXTs) are a subclass of supergiant high-mass X-ray binaries (SgXBs) mostly known for their peculiarly short X-ray outbursts, lasting a few hours at the most, 
and the much reduced average X-ray luminosity compared to the so-called classical SgXBs \citep[see, e.g.,][for a recent review]{walter2015}. 
In both classical SgXBs and SFXTs, the bulk of the X-ray emission is due to the accretion of the supergiant star wind 
onto a compact object, typically a neutron star (NS). The origin of the extreme X-ray variability of the SFXTs is still 
not well understood, but it is generally believed that the short outbursts are triggered by the presence of dense clumps 
in the stellar wind surrounding the NS \citep{zand05,walter07,negueruela06,oskinova2012,bozzo16}, while the reduction in the average 
luminosity can be ascribed to different mechanisms that inhibit the accretion for a large fraction of the time. 
The proposed mechanisms comprise either a magnetic and/or centrifugal gating \citep{grebenev07,bozzo08}, or the onset of a 
quasi-spherical settling accretion regime \citep[]{davies81,shakura12}. 

\src\ is the prototype of the SFXTs and was discovered in 2003 with \inte\ during a 2 hour long flare \citep{sunyaev03}. 
Since then, the source has been showing the most extreme X-ray variability among all other objects of the same class. 
It is characterized by an orbital period of 4.9~days \citep{clark09}, one of the shortest measured among the SFXTs, 
and a quiescent X-ray luminosity that can be as several times $\sim$10$^{32}$~\ergsec \citep{zand05}. Bright outbursts from 
\src\ were observed on many occasions by different instruments \citep[see, e.g.,][and references therein]{sguera06,rampy09,romano11}. 
So far, the most luminous event was caught by \swift\ in 2014 reaching approximately 3$\times$10$^{38}$~\ergsec in the 0.5-10~keV energy range 
\citep[assuming a source distance of 3.5~kpc;][]{pellizza06,rahoui08}. In the paper reporting the discovery of such intense 
X-ray emission \citep{romano15b}, the authors suggested that a temporary accretion disk could have formed around the NS, as the high 
luminosity recorded is virtually impossible to be reached in a wind-fed system. In the same paper, evidence of a 
possible pulsation at 11.6~s was reported, but never confirmed \citep[as well as the previous hint at 71~s;][]{drave12}.  
A likely cyclotron line at 17~keV has been discovered in a \nustar\ observation carried out in 2013 
\citep{bhalerao15}, making \src\ the first SFXT for which a direct measurement of the magnetic field is 
available ($B\simeq1.5\times10^{12}$~G). The properties of the supergiant companion hosted in \src\ were recently 
investigated in depth by \citet{gimenez16}, confirming it to be a O9I supergiant.   

In this paper, we report on a 150~ks-long observational campaign performed in the direction of \src\ using \xmm\ and \nustar\ simultaneously.\ 
During these observations, the source remained in a very low quiescent state for most of the time, and then, toward the end of the observations, 
it underwent a bright outburst comprising three distinct short flares lasting in total approximately 7~ks. We took advantage of the 
high statistics and good energy resolution of the instruments on-board \xmm\ and \nustar\ to investigate the properties of the source's  
X-ray emission during the quiescence and outburst periods. We also report on the results of the \swift\,/XRT monitoring performed  
for approximately one week around the time of the \xmm\ and \nustar\ observations and covering more than one orbital revolution of the system. 
We summarize the data analysis techniques in Sect.~\ref{sec:data} and present the 
results in Sect.~\ref{sec:pheno} and \ref{sec:physical}. The results of several optical and infrared observations   
carried out during the source quiescent period caught by \xmm\ and \nustar\ are summarized in Sect.~\ref{sec:others}. 
Our discussion and conclusions are provided in Sect.~\ref{sec:discussion}.

\section{X-ray data analysis}
\label{sec:data}

\subsection{\xmm\ }
\label{sec:xmm}

\begin{figure}
 \centering
   \begin{tikzpicture}[]
     \node at (1,1) {\includegraphics[width=6.5cm,angle=-90]{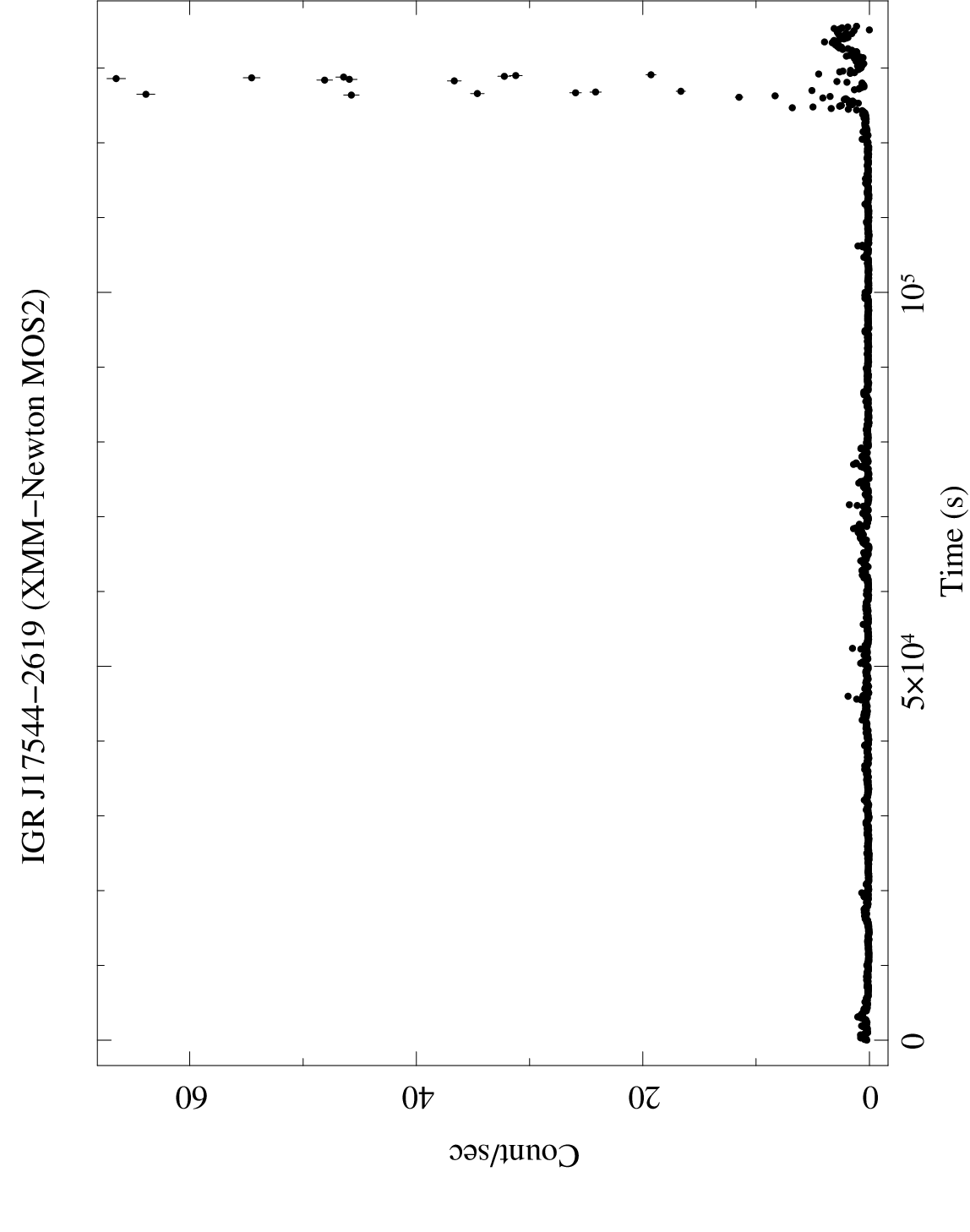}};
     \node at (1.05,1.35) {\includegraphics[scale=0.23,angle=-90]{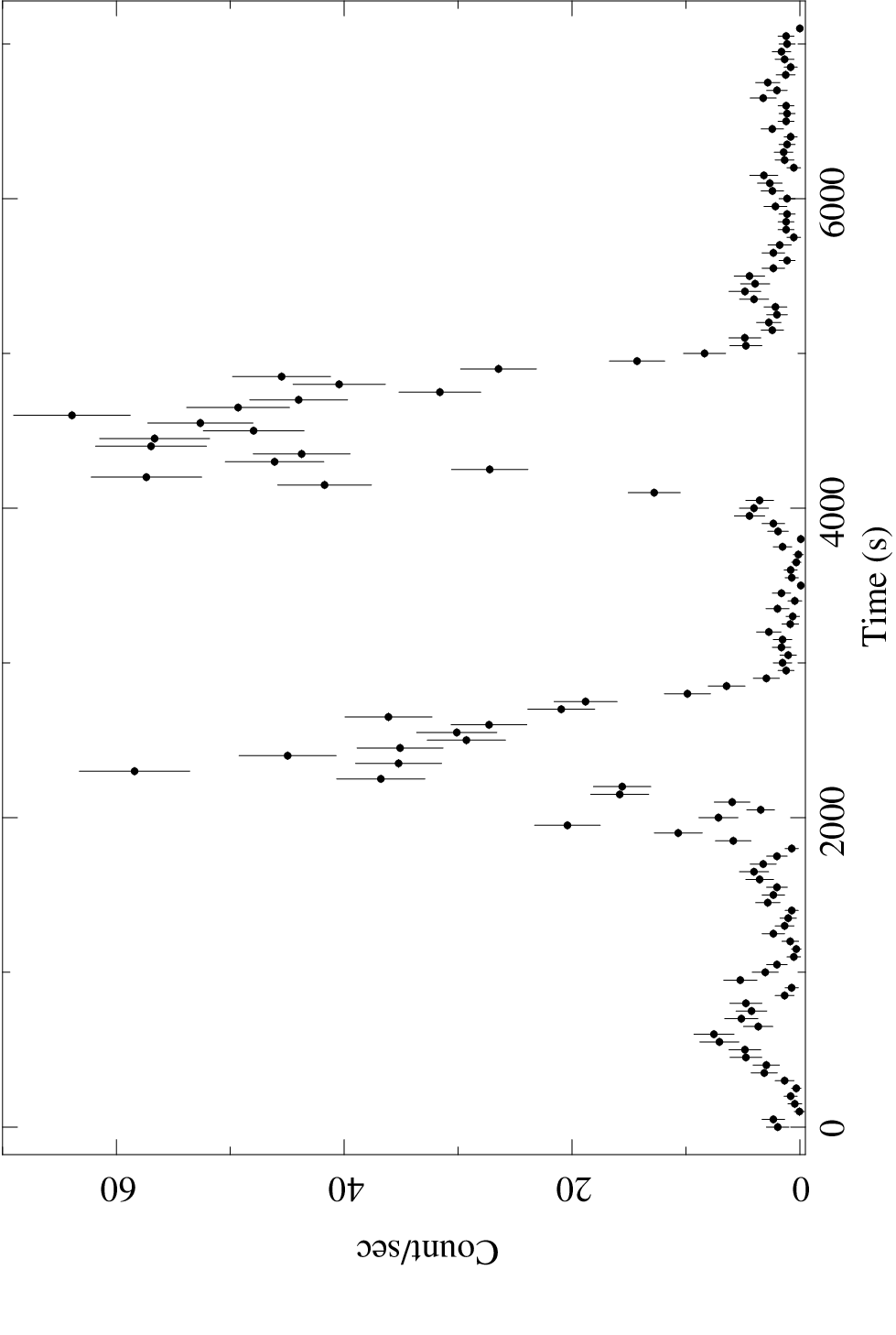}};
       \end{tikzpicture}
        \begin{tikzpicture}[]
     \node at (2,2) {\includegraphics[width=6.2cm,angle=-90]{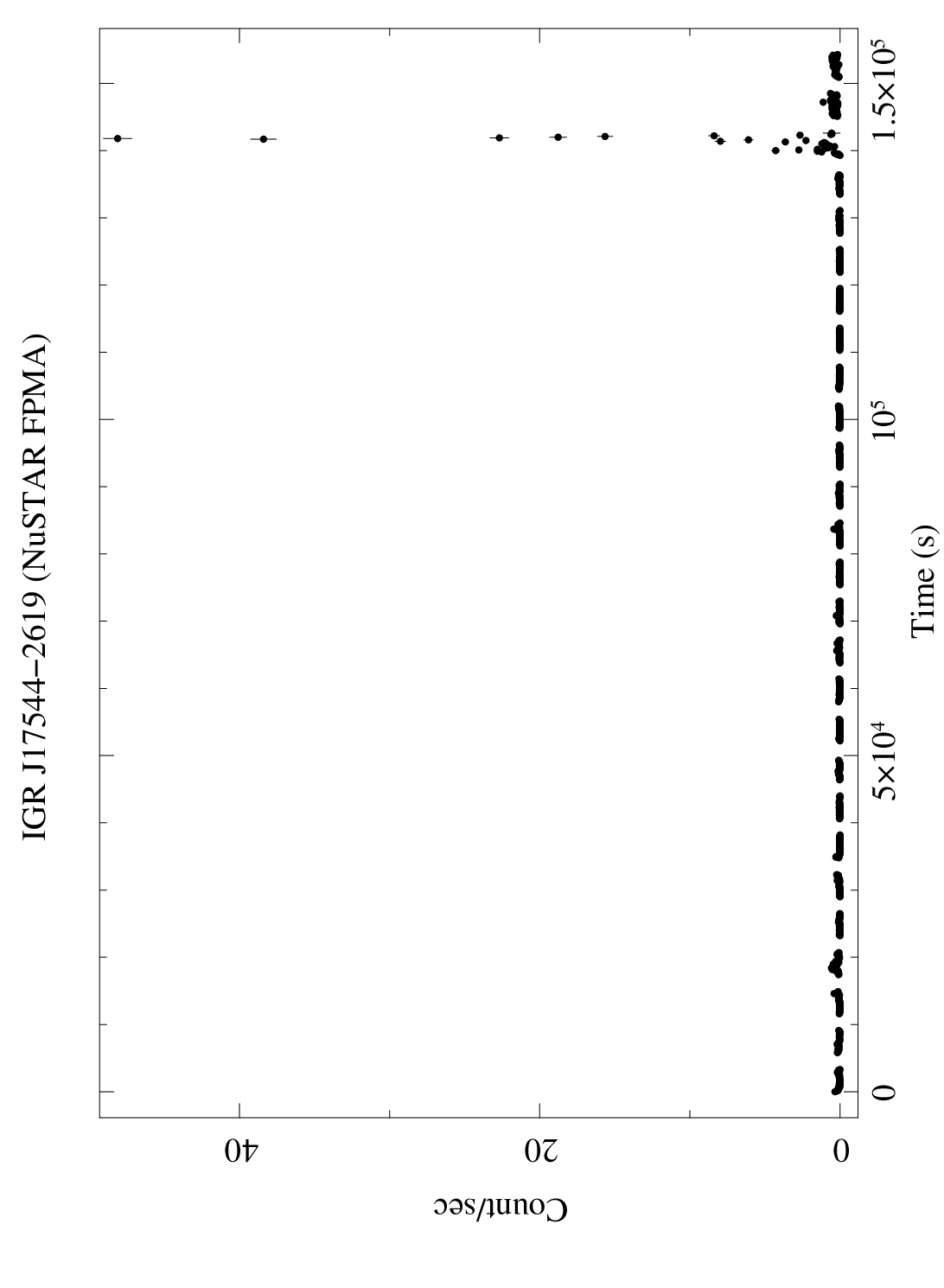}};
     \node at (2.05,2.35) {\includegraphics[scale=0.23,angle=-90]{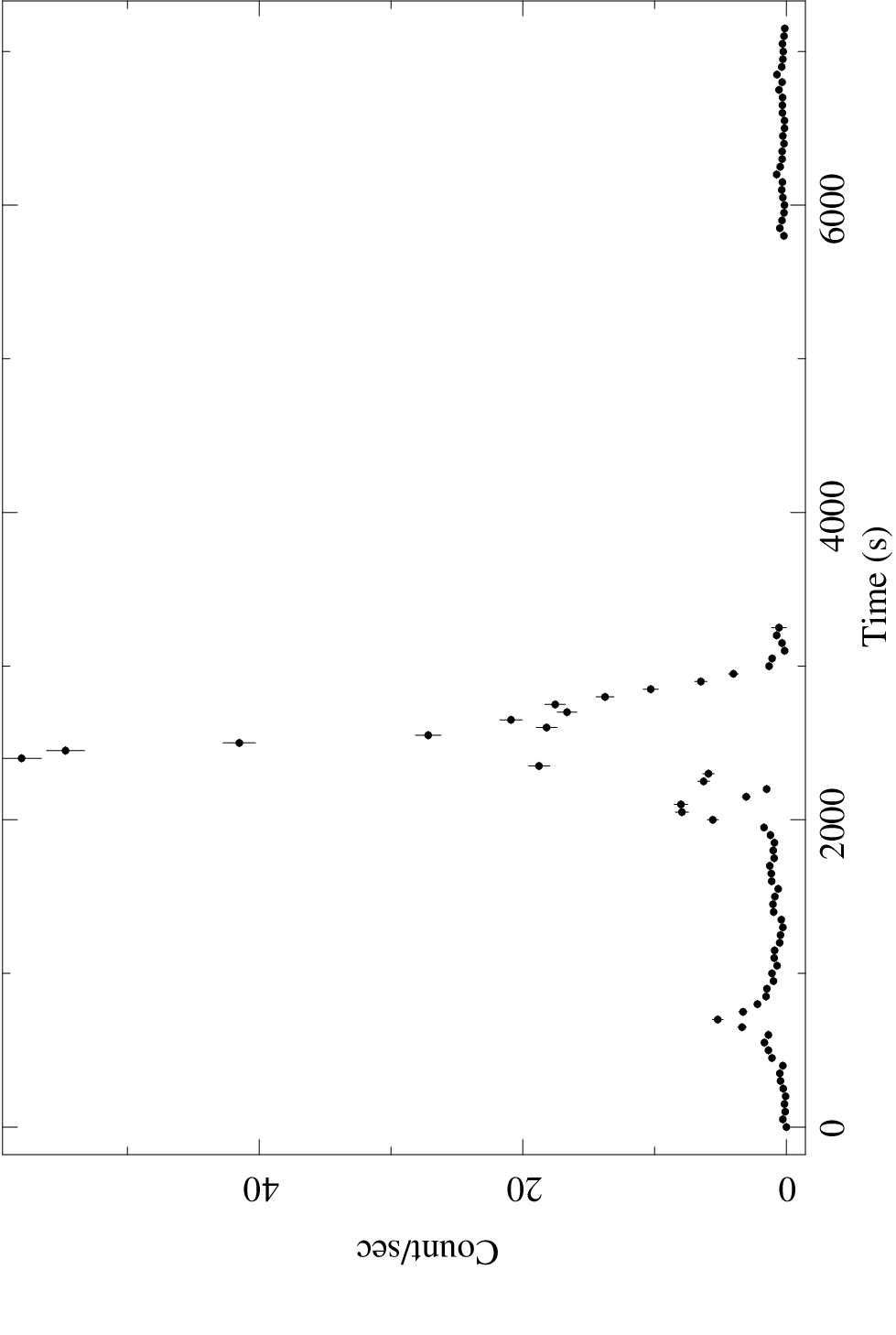}};
  \end{tikzpicture}
  \caption{{\it Top}: The \xmm\ MOS2 lightcurve of \src\ in the 0.5-10~keV energy band. The lightcurve has not been 
  filtered for the high flaring background time intervals. The insert shows 
  a zoom in the outburst of the source occurring towards the end of the observation and comprised of three distinct flares. 
  The time bin of the main lightcurve is 100~s, while for the lightcurve in the insert we used a time bin of 50~s. 
  The start time of the main lightcurve is 2015 March 20 at 06:02:34 UTC (57101.2518~MJD), while the start time of 
  the zoomed lightcurve in the insert is 2015 March 21 at 15:48:44 UTC (57102.6588~MJD). {\it Bottom}: the \nustar\ FPMA 
  lightcurve in the 5-10~keV energy band. The time bin is 100~s and the start time is 2015 March 20 at 1:06 UTC (57101.0458~MJD). 
  The inset shows a zoom into the flaring part of the \nustar\ lightcurve (in this case the start and the bin times are the same as 
  those of the inset in the top figure).}     
  \label{fig:lightcurve}
  \vspace{-0.5cm}
\end{figure}

The \xmm\ observation of \src\ began on 2015-03-20 05:00:31 UT and lasted until 2015-03-21 20:17:09 UT (OBSID: 0744600101; PI: E. Bozzo), 
resulting in a total exposure time of $\sim$141~ks (i.e., approximately 30\% of the source orbital period). 
The EPIC-pn and MOS1 cameras were operated in full frame, while the 
MOS2 was in timing mode. Data were also collected with the two grating instruments RGS1 and RGS2. 
All observation data files (ODFs) were processed using the \xmm\ Science Analysis System 
(SAS 15.0) following standard procedures\footnote{http://www.cosmos.esa.int/web/xmm-newton/sas-threads}. 
The observation was heavily affected by a flaring background during the first $\sim$120~ks. 
Removing the high background time interval resulted in an effective exposure time of 
approximately 60~ks for all the EPIC cameras and the two RGSs. 
The regions used for the extraction of the source spectra and lightcurves were chosen 
for all instruments to be centered on the best known position of \src, as reported by 
\citet{zand05}. The source displayed a large variability in X-rays (more than three orders of magnitudes, 
see Fig.~\ref{fig:lightcurve}), and thus the size and shape of the extraction regions 
for all instruments had to be carefully changed for different time intervals. 
We used a circular region during periods in which the source count-rate 
was $\lesssim0.5$~cts~s$^{-1}$ for the MOS1 and $\lesssim2.0$~cts~s$^{-1}$ for the 
pn\footnote{https://heasarc.gsfc.nasa.gov/docs/xmm/uhb/epicmode.html}. At higher count-rates, 
an annular region was used to avoid pile-up issues. The size of the inner hole of the annuls 
was changed from 50 to 450 pixels depending on the brightness of the source. The external 
radius of both the circular and annular region was also varied with this source intensity, ranging 
from 650 pixels during the quiescent period to 1000 pixels during the brightest part of the flares. 
For each pn and MOS1 spectra in which the correction for pile-up was needed, the removal of such effects 
was verified by comparing the results of the spectral fits with the MOS2 data. The latter, being in 
timing mode, was not affected by pile-up even during the peak of the flares (we cross-checked this finding by using 
the SAS {\sc epaplot} task). The background spectra and lightcurves were extracted for all EPIC cameras 
by using regions close to the position of \src\ but not contaminated by the source emission. All lightcurves were  
corrected for any remaining instrumental effect (e.g., vignetting) by using the {\sc epiclccorr} task. 
Throughout this publication we indicate uncertainties on all quantities at 90\% confidence level, unless stated otherwise. 

\begin{figure}
\centering
   \includegraphics[width=6cm,angle=-90]{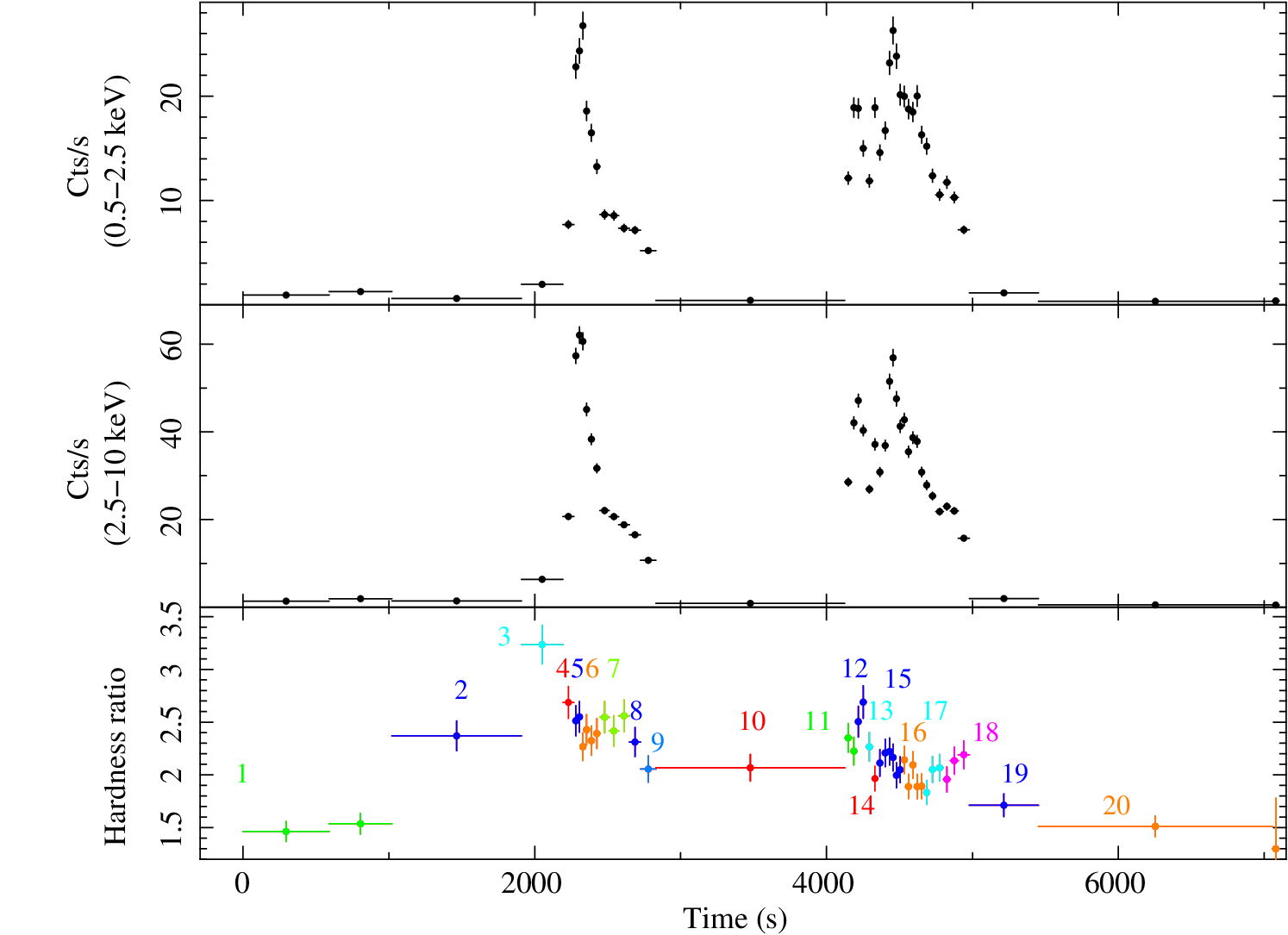}
  \includegraphics[width=6cm,angle=-90]{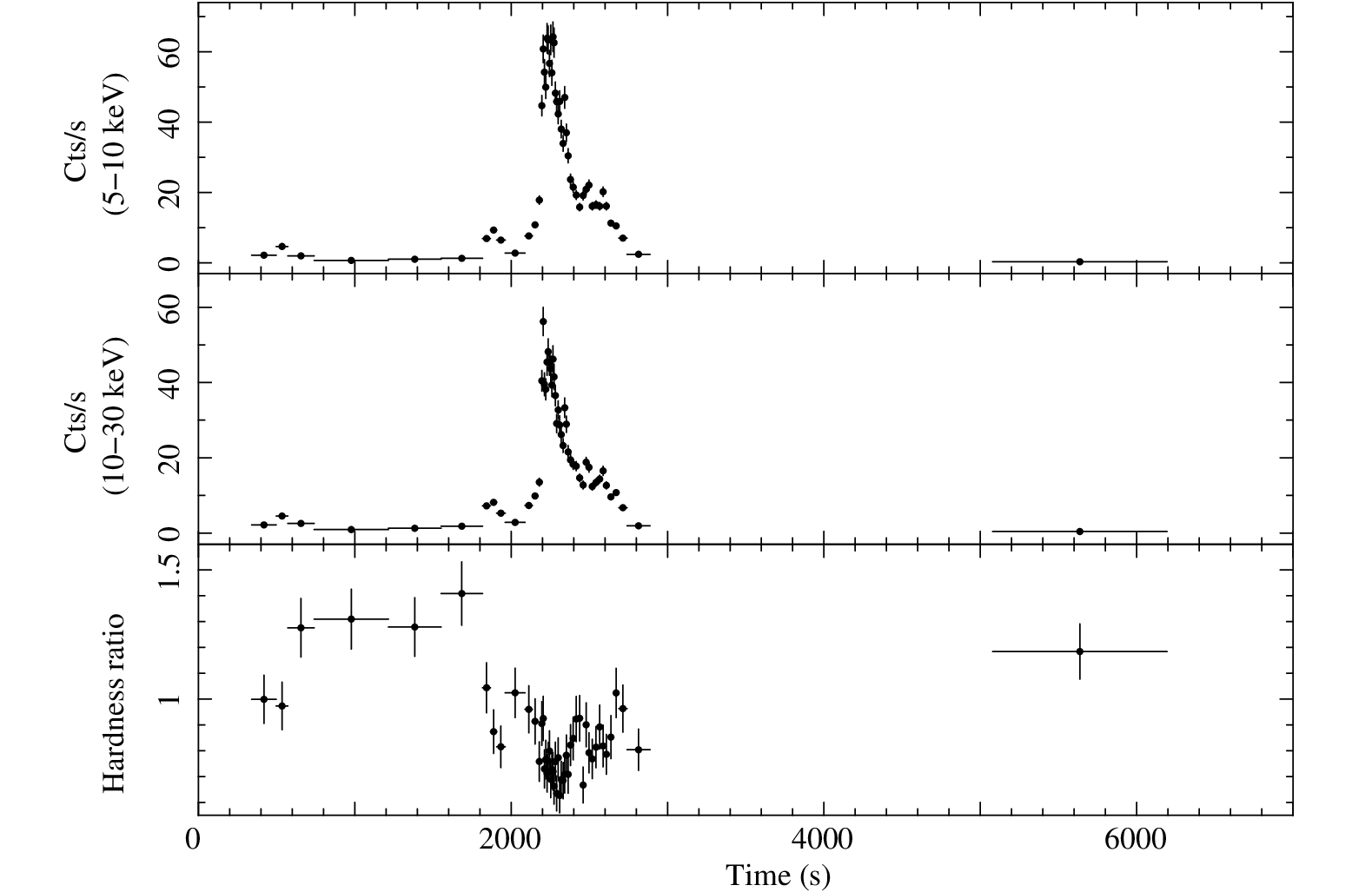}
  \caption{Lightcurves in different energy bands and the corresponding hardness ratio obtained from the 
  flaring part of the \xmm\ (top) and \nustar\ (bottom) observations. In both cases an adaptive rebinning has been used to 
  achieve  S/N$\gtrsim$15 in each bin. We do not show the quiescent portion of the lightcurve, as there, the 
  statistics were much lower than during flares and thus no meaningful HR resolved spectral analysis could be carried out. 
  Here the start time of the \nustar\ and \xmm\ lightcurves are the same. The third flare could not be 
  detected by \nustar\ due to visibility constraints. In the top figure, we highlighted the 20 intervals 
  in which the HR-resolved spectra analyzed in Sect.~\ref{sec:pheno} have been extracted with different colors.} 
  \label{fig:hr}
\end{figure}

We show the MOS2 lightcurve of the source in the 0.5-10~kev energy range in Fig.~\ref{fig:lightcurve}, 
as this one is not affected by pile-up.  
As anticipated in Sect.~\ref{sec:intro}, the source remained in a low emission state for the initial 
120~ks of the observation and then underwent a 7~ks long outburst, comprising three fast flares occurring 
approximately during the periastron passage (see details in Sect.~\ref{sec:discussion}).  
The first flare was much fainter than the following two (by a factor of 10-20). 
The average 0.5-10~keV count-rate of the source during the quiescent period in the MOS2 was of 
0.090$\pm$0.001~cts~s$^{-1}$, while the peak count-rate registered during one of the three flares 
achieved 82$\pm$12~cts~s$^{-1}$ when the lightcurve was binned at 10~s. 
In Fig.~\ref{fig:hr} we also show  the hardness ratio (HR) of two MOS2 lightcurves extracted in the 
0.5-2.5~keV and 2.5-10~keV energy bands. An adaptive rebinning has been used, following the same technique adopted in a number of 
our previous papers \citep[see, e.g.,][]{bozzo13b}, to achieve 
a minimum signal-to-noise ratio (S/N) of 15 at each point. 

Evidence for coherent and quasi-coherent oscillations in the \xmm\ data were searched for using event files
from the source where the arrival times of all photons were barycentre-corrected. 
No significant detection could be found in the all accessible frequency range ($\sim$10$^{-5}$-300\,Hz).

\subsection{\nustar\ DATA}
\label{sec:nustar}

IGR~J17544--2619 was observed by \nustar\ from 2015 March 20 at 
00:51:07 to March 21 at 20:01:07 (UTC; PI: Bhalerao). The Target of Opportunity observation 
(OBSID~90001005002; PI: Bhalerao) was triggered in order to obtain as much 
simultaneous data as possible with the scheduled \xmm\ observation (see Sect.~\ref{sec:xmm}). 
After having applied all the good time intervals (GTI) to the \nustar\ data  accounting for the Earth occultation and 
the South Atlantic Anomaly passages, we obtained an effective exposure time of 61.3~ks and 62.8~ks for the  
Focal Plane Modules A and B, respectively. The data were processed using {\sc nustardas v1.5.1}, and 
{\sc CALDB} dated 2015 September 4. The source photons were extracted from a 60~arcsec circle centred on the source, 
while the background was evaluated using polygonal extraction regions on the same chip. In NuSTAR FPMB, the source region 
was contaminated by stray light. We selected a background region with the same level of contamination to ensure that this stray light 
did not affect our results. The average source count-rate recorded for most of the observation was  
approximately 0.1-0.2~cts~s$^{-1}$ (3-80~keV energy band), yielding nearly 100\% live time. During the source outburst 
at the end of the observation, a count-rate as high as $\sim50$~counts/sec was measured. 

We show in Fig.~\ref{fig:lightcurve} the entire \nustar\ lightcurve of \src\ as observed by the FPMA in the 5-10~keV energy 
band, while in Fig.~\ref{fig:hr} we report a zoom in the \nustar\ lightcurves extracted in the 3-10~keV and 10-30~keV, together with the 
corresponding HR. In this plot, an adaptive rebinning has been used to achieve S/N$\gtrsim$15 in each time bin as for the 
\xmm\ case (see Sect.~\ref{sec:xmm}). 
\nustar\ observed only two of the three flares displayed by the source during the outburst. The third flare could not be observed 
due to the satellite visibility constraints. 

Due to the lack of any significant detection in the \xmm\ data (see Sect.~\ref{sec:xmm}) and the 
known issues affecting X-ray timing analyses with \nustar\ \citep[see][and references therein]{bachetti15}, we did not perform 
detailed searches for coherent and quasi-coherent modulations of the events recorded by the FPMA and FPMB.

\subsection{\swift\ DATA}
\label{sec:swift}

\src\ was observed by {\it Swift} as a ToO campaign (PI: Romano) 
aimed at monitoring the general flux level of the source around 
the {\it XMM-Newton} observation, with six daily observations, each 5\,ks long, 
starting on 2015 March 18. As the orbital period of 
the source is approximately 4.9~days (see Sect.~\ref{sec:intro}), the XRT \citep{burrows05} and UVOT \citep{roming05} 
data covered slightly more than an entire orbital revolution of \src.\ The complete 
log of the XRT observations is provided in Table~\ref{xrtobslog}, while for UVOT we summarize all 
relevant information in Table~\ref{tab:uvot}. 

The XRT data were processed and analyzed using the standard software ({\sc ftools} v6.16),  
calibration files (CALDB 20140709), and methods. 
All data were processed and filtered with {\sc xrtpipeline} (v0.13.1).  
The source remained at a level of a few $\times 10^{-2}$\, counts\,s$^{-1}$ throughout the
campaign, and only flared up to 0.17~cts~s$^{-1}$ on March 22 (obs.~00035156174). 
As the source was never affected by pile-up, all events were accumulated within a circular region 
with a radius of 20 pixels (where 1 pixel corresponds to $\sim2\farcs36$), while 
background events were accumulated from an annular source-free region
with inner/outer radii of 70/100 pixels centered on the source.  
When no detection was achieved in one observation, we calculated the corresponding 3\,$\sigma$ upper limit 
by using the {\sc sosta} and {\sc uplimit} tools available within {\sc XIMAGE}, together with  
the Bayesian method for low count experiments  \citep{kraft91}. 
The XRT lightcurves were corrected for point spread function losses, vignetting, and 
were background subtracted. 
\begin{table} 
{\fontsize{9}{4}\selectfont 
 \begin{center}         
 \caption{Log of all {\it Swift}/XRT observations used in the present paper.}         
 \label{xrtobslog} 
 \bgroup        
 \def\arraystretch{1.8}
 \begin{tabular}{@{}lllll@{}} 
 \hline 
 \hline 
 \noalign{\smallskip} 
 Sequence   & Obs/Mode  & Start time  & End time & Exposure   \\ 
                  &      & (MJD)  & (MJD)  &(s)     \\
  \noalign{\smallskip} 
 \hline 
 \noalign{\smallskip} 
00035056169     &       XRT/PC  &       57099.2711      &       57099.4715      &       3395    \\
00035056170     &       XRT/PC  &       57100.6156      &       57100.8097      &       3896    \\
00035056171     &       XRT/PC  &       57101.0023      &       57101.2735      &       878     \\
00033707001     &       XRT/PC  &       57101.0038      &       57101.2853      &       3962    \\
00033707002     &       XRT/PC  &       57101.1454      &       57101.1569      &       980     \\
00035056173     &       XRT/PC  &       57102.0620      &       57102.2798      &       4874    \\
00035056174     &       XRT/PC  &       57103.5947      &       57103.8041      &       4969    \\
00035056175     &       XRT/PC  &       57104.5921      &       57104.7381      &       2841    \\
00035056176     &       XRT/PC  &       57107.7829      &       57107.8582      &       2061    \\
  \noalign{\smallskip}
  \hline
  \end{tabular}
  }  
  \end{center}
  %\begin{list}{}{} 
  %\item[$^{\mathrm{a}}$] 
  %\end{list} 
  }
  \end{table}
\begin{figure}
\centering
   \includegraphics[width=6.8cm,angle=-90]{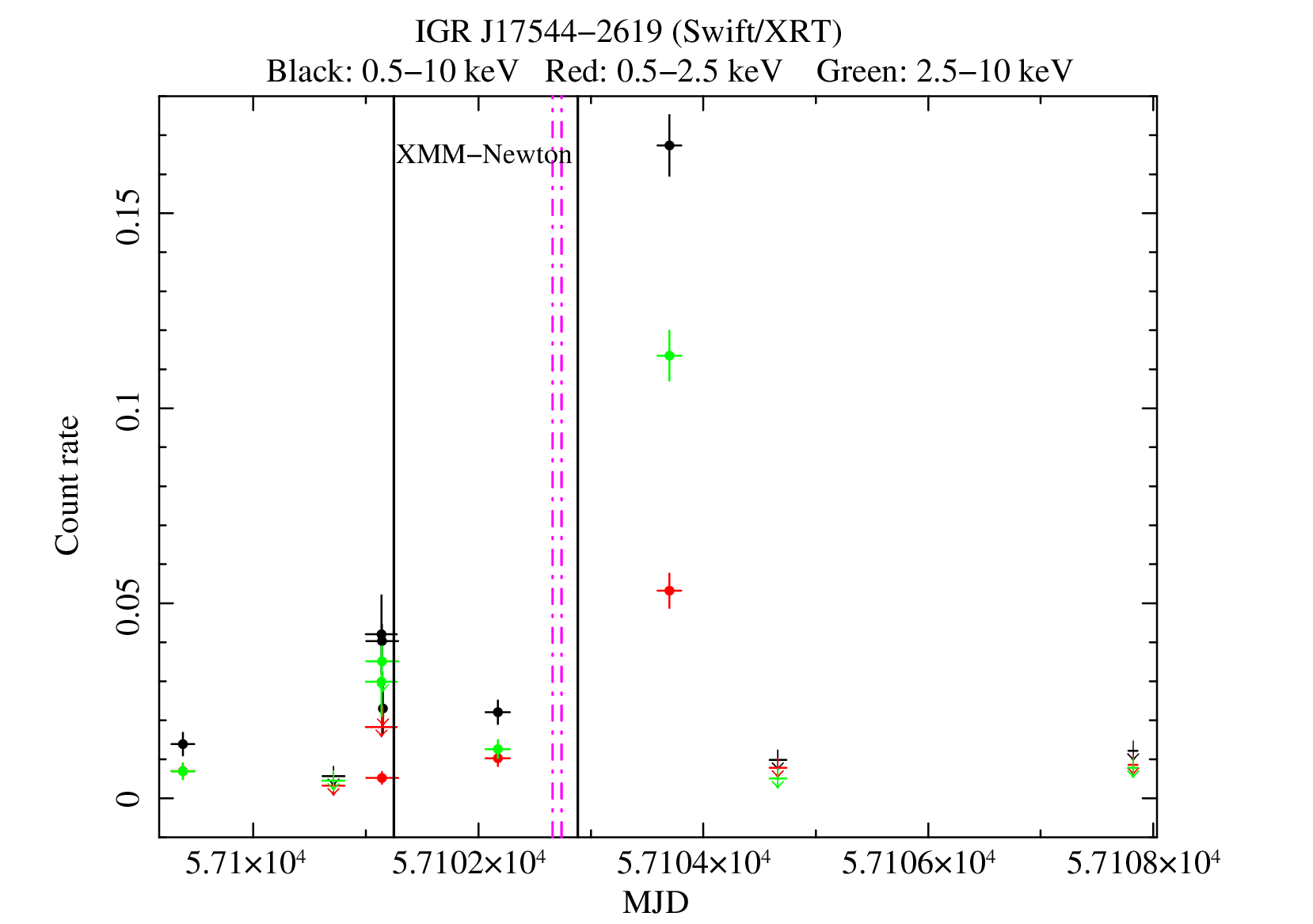}
  \caption{Lightcurves obtained from the \swift\,/XRT monitoring campaign performed a few days before and after the \xmm\ 
  and \nustar\ observations of \src\ (we mark in black the lightcurve in the 0.5-10~keV energy range, 
  in red the lightcurve in the 0.5-2.5~keV energy range, and in green the lightcurve in the 2.5-10~keV energy range). 
  The time is measured from the 57102~MJD, as in Fig.~\ref{fig:lightcurve}. 
  The faint flare recorded by XRT, which is visible above when the source count-rate reaches approximately 0.17~cts~s$^{-1}$, occurred on 
  57103.7~MJD, that is, roughly one day after the onset of the much brighter outburst observed by \xmm\ and \nustar.\ 
  The downward arrows indicate 3$\sigma$ upper limits on the source count-rate when \src\ was not detected in the 
  corresponding XRT observation. We also marked with vertical solid lines the time interval of the \xmm\ observation (the \nustar\ 
  observation is nearly simultaneous). The vertical, dashed magenta lines indicate the time interval of the 7~ks outburst detected by \xmm\ 
  and \nustar\ (see Sect.~\ref{sec:xmm} and \ref{sec:nustar}).} 
  \label{fig:swift}
\end{figure}

The XRT data showed that the source remained in a relatively low X-ray emission state during the entire monitoring campaign. 
Only a short flare was recorded, reaching approximately 0.17~cts~s$^{-1}$ in the energy band of the instrument. Excluding this flare, 
the average count-rate recorded by XRT from the source was approximately 0.025$\pm$0.003~cts~s$^{-1}$ in the 0.5-10~keV energy range. 
This corresponds to a flux of 2.7$\times$10$^{-12}$~\eflux when an absorbed power-law model with 
$N_{\rm H}$=2$\times$10$^{22}$~cm$^{-2}$ and $\Gamma$=1 is used for the conversion (see Sect.~\ref{sec:pheno}).   
Due to the relatively low statistics of the XRT observations, no useful timing and spectral analysis of these 
data could be carried out. 

UVOT observed \src\ simultaneously with the XRT using different 
filters (B, M2, U, V, W1, and W2) in different observations in order to provide the broadest 
wavelength coverage possible. The data analysis was performed
using the {\sc uvotimsum} and {\sc uvotsource} tasks included in the {\sc ftools} 
software. The latter task calculates the magnitude through aperture
photometry within a circular region and applies specific corrections
due to the detector characteristics. We provide a summary of all results in 
Table~\ref{tab:uvot}, where the magnitudes have been computed for each filter by combining 
data with the same filter in different observations. The reported magnitudes are on
the UVOT photometric system described in \citet{uvot11} and
are not corrected for Galactic extinction. We did not find significant differences in the 
estimated magnitudes of the source in all filters compared to values obtained during 
previous monitoring campaigns \citep[and compatible with expected values for the  
supergiant star hosted in \src;][]{romano11}. 
\begin{table}
\tiny
\caption{Summary of UVOT results. The indicated uncertainties on the observational times 
correspond to the time coverage of the combined images obtained with the same filter in different 
observations. 
\label{tab:uvot}}
\begin{center}
\begin{tabular}{@{}lllll@{}} 
\hline
MJD$^a$ & Filter & Magnitude \\
57103.5639$\pm$4.2257      & B      & 14.53$\pm$0.03      \\
57104.2429$\pm$3.5599      & M2     & 20.29$\pm$0.15      \\
57103.5623$\pm$4.2256      & U      & 15.19$\pm$0.03       \\
57104.2397$\pm$3.5582      & V      & 12.81$\pm$0.02       \\
57103.5647$\pm$4.2936      & W1     & 16.63$\pm$0.04       \\
57103.5680$\pm$4.2281      & W2     & 17.92$\pm$0.05       \\
\hline
\end{tabular}
\end{center}
\end{table}

\section{Spectral analysis with phenomenological models}
\label{sec:pheno}

We first extracted the \xmm\ and \nustar\ spectra averaged during the entire observation and only during the quiescent period. 
When averaged over the entire observation, both the MOS1 and the pn data were heavily 
piled-up and the comparison with the MOS2 data revealed that it was not possible to find a satisfactorily correction 
for this issue to obtain compatible results for all cameras simultaneously. For this reason, we only used the MOS2, 
RGS, and \nustar\ data to analyze the properties of the source X-ray emission averaged over the entire available exposure time. 
During quiescence, the source emission was too faint to obtain meaningful RGS spectra, and thus we did not 
include the corresponding data in the combined \xmm\ and \nustar\ fit.  

\begin{figure}
\centering
   \includegraphics[width=6.8cm,angle=-90]{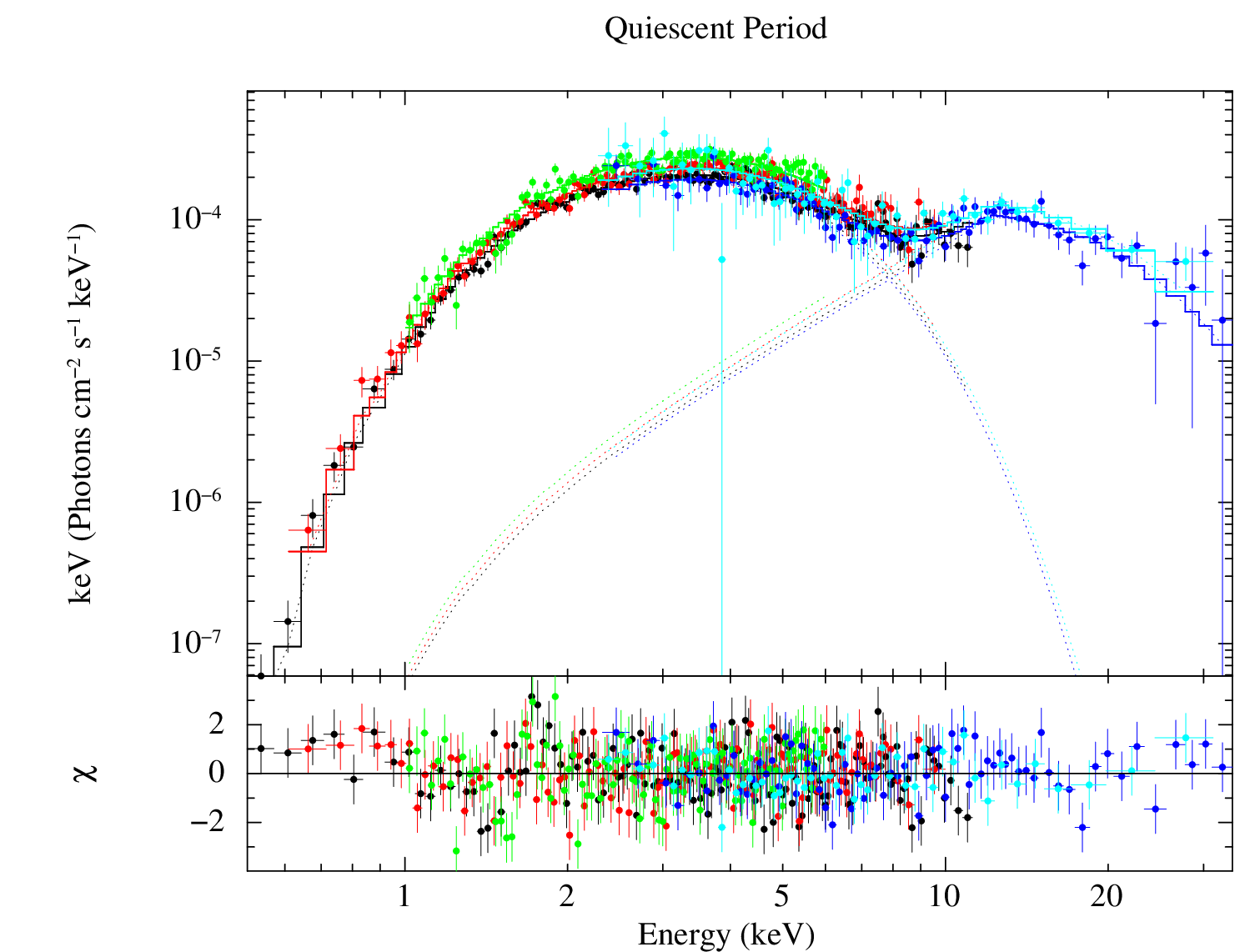}
   \includegraphics[width=6.8cm,angle=-90]{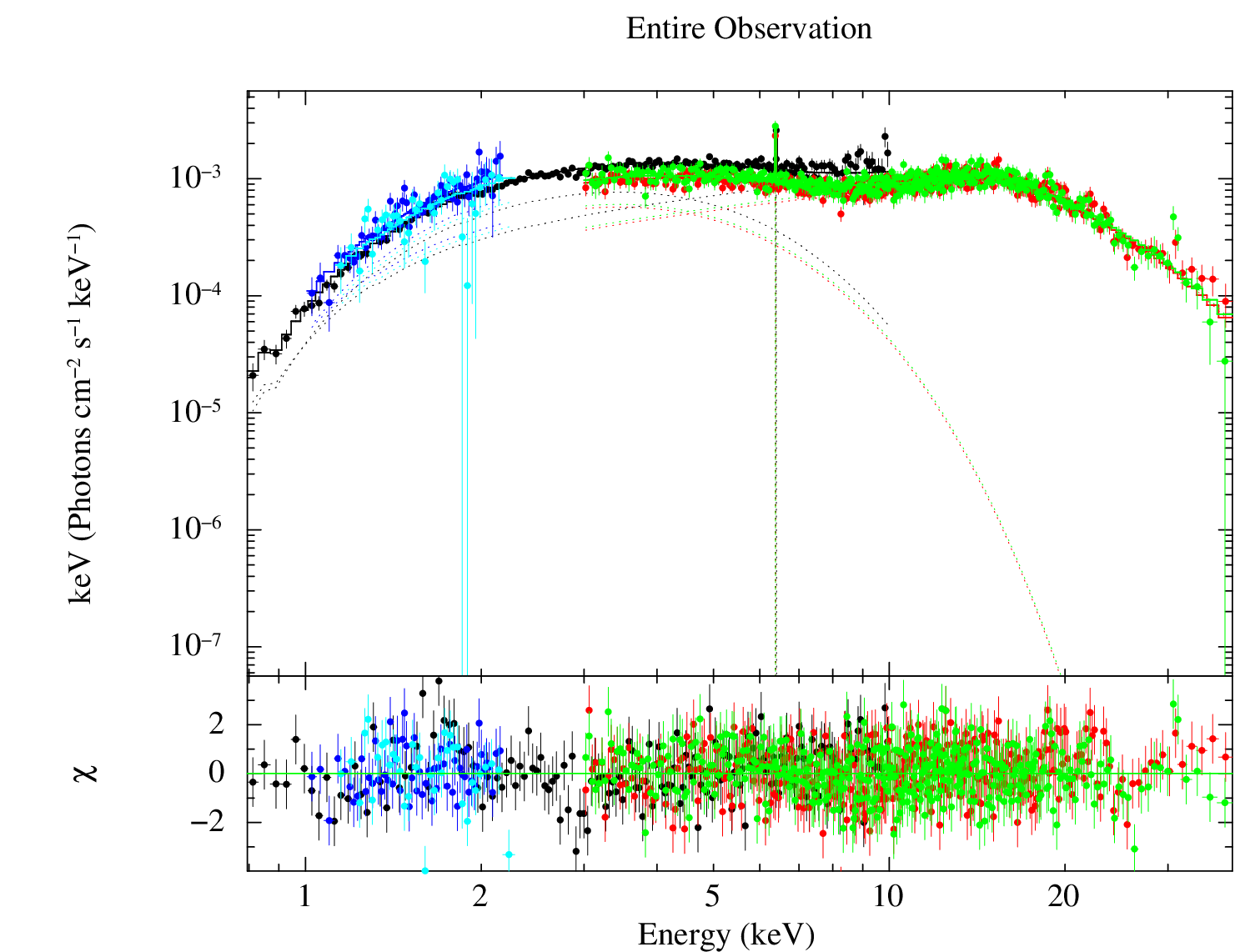}
  \caption{{\it Top}: the combined \xmm\ and \nustar\ unfolded spectra extracted during the 
  first $\sim$120~ks of the observations, when the source was in a quiescent state. The EPIC-pn is in 
  black, the MOS2 in red, the MOS1 in green, the FPMA in blue, and the FPMB in cyan. The best fit 
  model is obtained by using a combination of an absorbed blackbody plus a power-law with a high energy 
  cut-off (see Sect.~\ref{sec:pheno} for details). The residuals from the best fit are shown in the bottom panel. {\it Bottom}: 
  same as above but for the spectra accumulated during the entire observational period. The MOS2 is in 
  black, the FPMA in red, the FPMB in green, the RGS1 in blue, and the RGS2 in cyan. The same model as above 
  plus a weak iron emission line at $\sim$6.4~keV, has been used to obtain the best fit. 
  Residuals from this fit are shown in the bottom panel.}  
  \label{fig:spectra}
\end{figure}

\begin{figure*}
\centering
   \includegraphics[width=3.6cm,angle=-90]{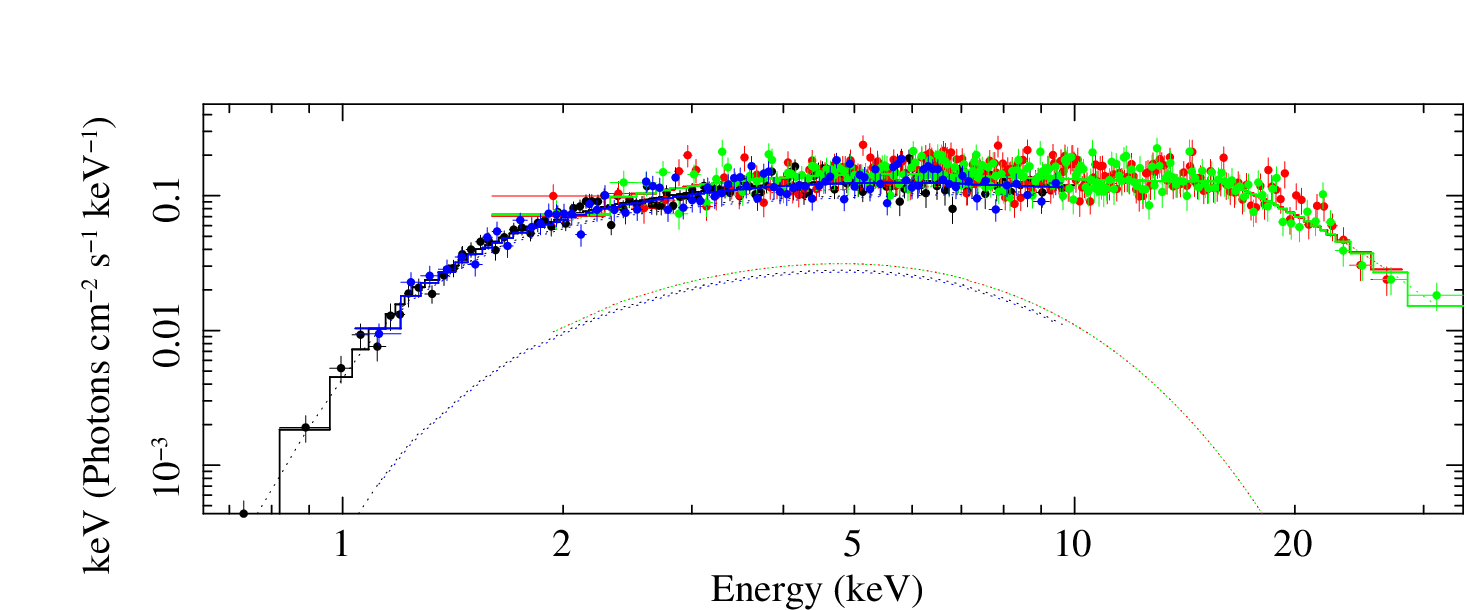}
   \includegraphics[width=3.6cm,angle=-90]{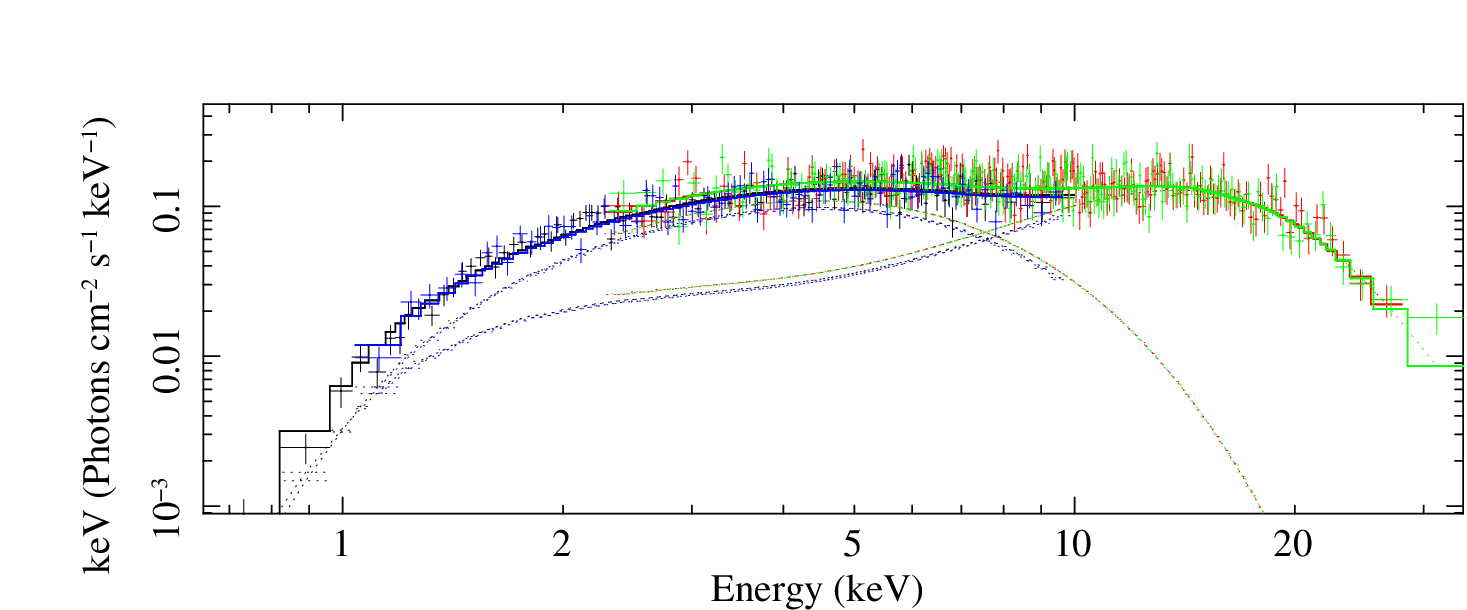}
   \includegraphics[width=3.6cm,angle=-90]{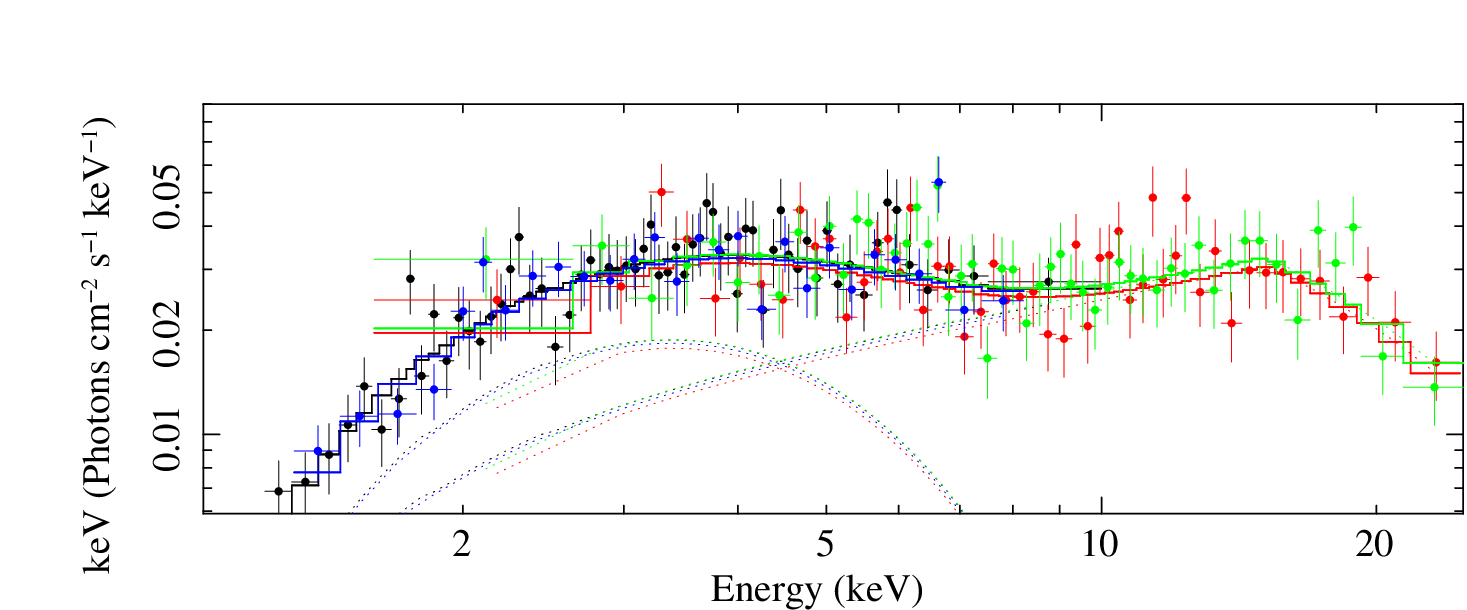}
   \includegraphics[width=3.6cm,angle=-90]{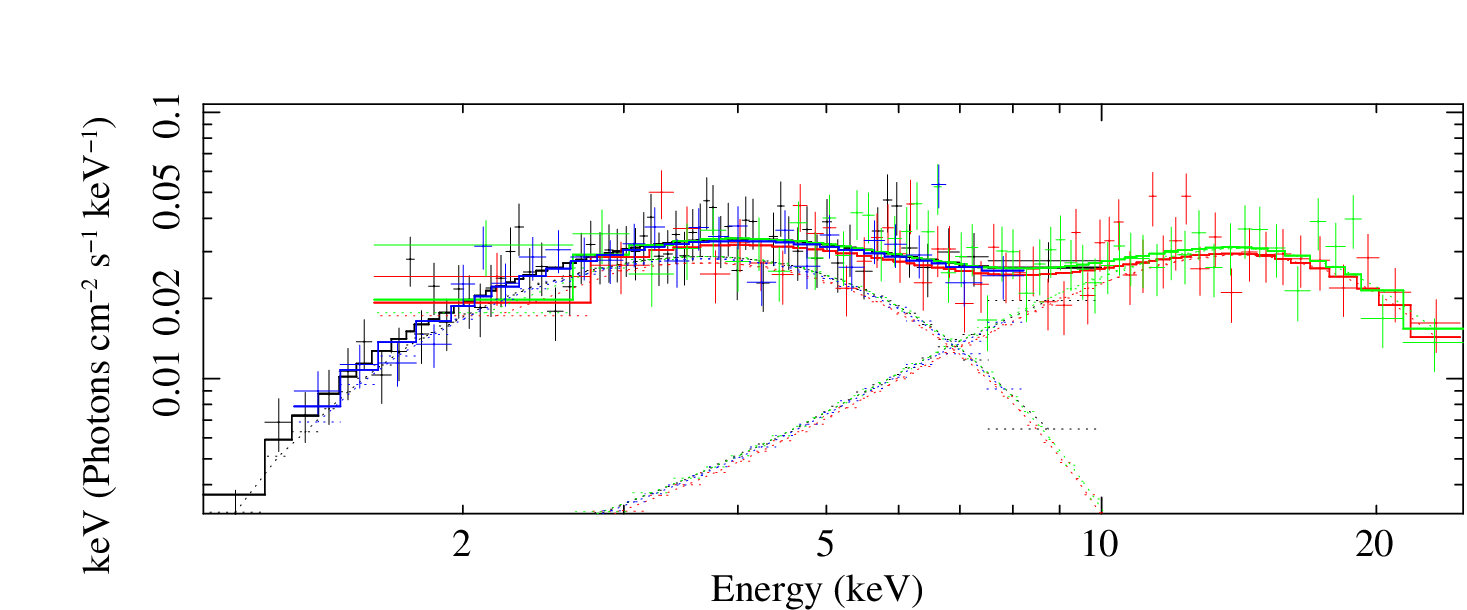}
  \caption{An example of combined \xmm\ and \nustar\ spectra during the decay from the second flare, as indicated in Fig.~\ref{fig:hr}. 
  We chose the spectra from the time interval 6 (top figures) and 9 (bottom figures) as examples to show how the source X-ray spectral 
  emission evolves during the flare. For the figures on the left, the non-thermal component is described through the usage of a phenomenological 
  {\sc highecut*pow} model in {\sc xspec}. For the figures on the right, the non-thermal component is fit by using the physical BW model 
  (see Sect.~\ref{sec:physical}). In both cases, it is evident that the non-thermal component dominates the source high-energy emission 
  at higher fluxes (e.g., during the time interval used to extract the spectrum 6). The thermal component provides an increasingly 
  important contribution to the overall emission during the decay from the flare (e.g., during the time interval used to extract the 
  spectrum 9). The relative contribution of the thermal and non-thermal components is different in the fits performed with the 
  phenomenological and physical spectral models, but the overall picture used to interpret the spectral change is qualitatively similar.}   
  \label{fig:otherspectra}
\end{figure*}

We adopted several different phenomenological models to describe the broad-band spectra of \src.\ Due to the  
complex spectral energy distribution in the 0.5-40~keV energy range 
(see, e.g., Fig.~\ref{fig:spectra} and \ref{fig:otherspectra}), we could only achieve acceptable results by using 
a combination of an absorbed thermal blackbody ({\sc bbodyrad} in {\sc xspec}) plus a 
power-law with a high-energy cut-off ({\sc highecut*pow} in {\sc xspec}). This model has often been 
used to describe the X-ray emission of highly magnetized accreting pulsars  
\citep[see, e.g.,][]{white95xrb, coburn2001}. The thermal component is most likely originating from  
the NS surface, while the power-law might originate from Comptonization 
within its accretion column or other non-thermal processes. A weak iron K$_\alpha$ line at 6.4~keV was detected in the spectra averaged 
over the entire observational period. This is likely produced by fluorescence due to the X-ray illumination 
of the stellar wind material surrounding the compact object hosted in \src,\ as the measured equivalent width 
is relatively small and comparable to that observed in a large number of wind-fed HMXBs 
\citep[see, e.g.,][and references therein]{bozzo08,torrejon10,manousakis11,gimenez15}.  
We summarize all results obtained with the phenomenological model in Table~\ref{tab:fits} and show in Fig.~\ref{fig:spectra} 
the unfolded combined \xmm\ + \nustar\ spectra corresponding to the entire observation and the quiescent period only. 
\begin{figure}
\centering
   \includegraphics[width=6.8cm,angle=-90]{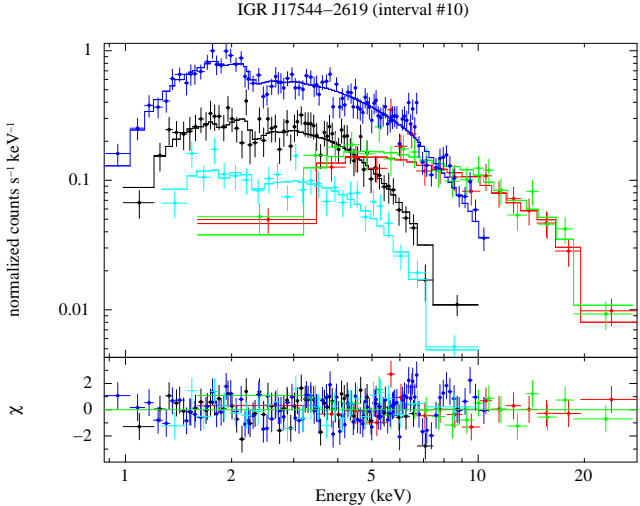}
  \caption{The \xmm\ and \nustar\ spectra extracted during the time interval 10. This is the only interval in which 
  a spectral feature is observed around $\sim$7.2~keV. The best fit model to describe the continuum in this case is the same 
  mentioned in Fig.~\ref{fig:otherspectra}. The residuals from the best fit are shown in the bottom panel and no component was added to 
  take the presence of the feature into account in order to highlight its significance.}  
  \label{fig:spectra10}
\end{figure}
\begin{figure}
\centering
   \includegraphics[width=7.6cm,angle=-90]{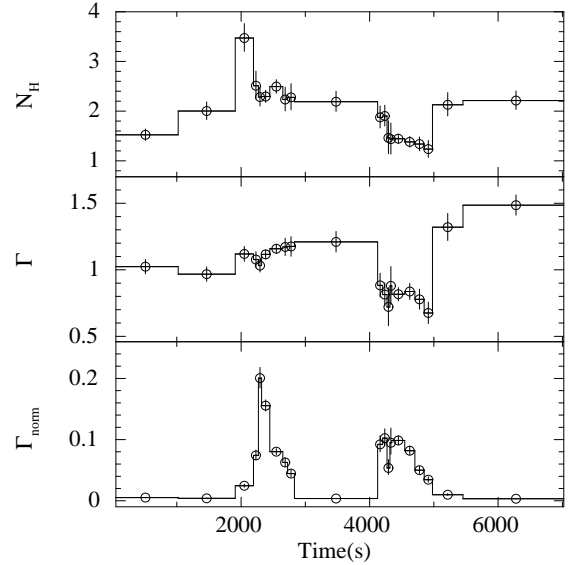}
  \caption{Results obtained when the time-resolved \xmm\ data alone are fitted with a simple absorbed power-law model. From the top to the bottom 
  panel we show the absorption column density, the power-law photon index, and its 
  normalization. As discussed in the text, it is evident that the most prominent spectral changes took place during the second flare (the first one 
  being much fainter than the other two and occurring in this figure at approximately $t$=500~s). The uncertainties on the $x$ axis correspond to the 
  integration times of the spectra 1-20 in Fig.~\ref{fig:hr}.}  
  \label{fig:xmmtimeresolved}
\end{figure} 

In order to study possible spectral variations that could give rise to the changes in the HR visible in Fig.~\ref{fig:hr}, 
we carried out a hardness ratio resolved spectral analysis of the outburst period. We did not carry out the same 
analysis during the quiescent period due to the much-reduced statistics of the data. We selected 20 different intervals, as shown 
in the upper panels of Fig.~\ref{fig:hr} and fit all of them with the same two-component model described above. As the third flare was not observed by 
\nustar\ due to the satellite visibility constraints, no high-energy coverage ($>$10~keV) was available for the spectra extracted 
during the time intervals 11-19. These time intervals are thus not included among all the broad-band results presented in 
Table~\ref{tab:fits}. 

From the values reported in the table we found that there was a significant increase in the absorption 
column density local to the source immediately before the rise to the second flare in the outburst (the first one was too faint 
to reveal any significant spectral variability). The $N_{\rm H}$ dropped by a factor of $\sim$1.5 
when the source reached the peak of the flare and then remained virtually constant throughout the rest of the observation. 
The thermal component is observed to increase in temperature and in radius toward higher fluxes, where the Comptonization 
component clearly becomes dominant. In order to visually show these variations, we report in Fig.~\ref{fig:otherspectra} the 
case of the spectra extracted during the time interval 6, close to the peak of the second flare, and the time interval 
9, half way through the decay from this flare. 

A puzzling result is found during the time interval 10, which corresponds to 
the low count-rate period separating the second and the third flare. The spectra extracted during this interval (only partially 
covered by \nustar) are shown in Fig.~\ref{fig:spectra10}. As the source flux in this interval dropped significantly, the pn spectrum 
was not affected by pile-up. An evident absorption feature is detected at 7.2~keV. By adding a Gaussian absorption component 
to the spectrum, we measured a centroid energy of 7.21$\pm$0.14~keV and a width of 0.2$\pm$0.1~keV (the latter being 
compatible with a broadening due to the limited energy resolution of the pn). The estimated equivalent width of the line 
and the associated uncertainty is 0.23$_{-0.06}^{+0.11}$~keV, thus suggesting a detection significance $>$3$\sigma$.   
Adding the line to the fit of all spectra in the interval 10 (i.e., the combined fits of MOS1, MOS2, pn, FPMA, and FPMB) 
leads to a decrease of the $\chi^2_{\rm red}$ from  
$\chi^2_{\rm red}$/d.o.f.=0.93/214 (fit without the line) to $\chi^2_{\rm red}$/d.o.f.=0.86/211 (fit with the line included). 
This again suggests a detection significance at the level of $\gtrsim$3$\sigma$ according to the F-test included within 
{\sc xspec}. 

The lack of \nustar\ data for the time intervals 11-19 prevented an analysis of the 
remaining time-resolved spectra comparable to that discussed above. For each interval, the \xmm\ data alone could be reasonably well-fitted by using a simple, absorbed power-law 
model. We verified that using the two-component models considered above for the broad-band spectral fits on the time-resolved \xmm\ spectra alone 
always resulted in very poorly constrained parameters. In order to have a complete overview of the time-resolved spectral analysis 
achievable with \xmm,\ we thus report in Fig.~\ref{fig:xmmtimeresolved} a plot of all spectral parameters obtained by fitting the  
\xmm\ data (MOS2 plus MOS1 and pn, when available) in the time intervals 1-20 with a simple absorbed power-law model. 
From this plot, it is evident that most of the spectral changes occurred during the second flare, while for the third flare 
only minor variations in the absorption column density and power-law photon index were measured. We were thus lucky 
to have the second flare, rather than the third one, observed simultaneously with \nustar.\ 

We note that in neither the time-resolved nor the averaged spectra could we find a clear indication for the presence of the cyclotron line 
at $\sim$17~keV reported previously by \citet{bhalerao15}. If a cyclotron line is added to the phenomenological model used in this section 
to fit the broad-band spectra of \src,\ the centroid energy of this feature is moved by {\sc xspec} approximately 8~keV and its width becomes as large 
as 2-3~keV. We did not consider this a reasonable model because a feature at 8~keV could not be the fundamental energy of the previously detected 
cyclotron lines from the source at 17 and 33~keV, and, additionally, the feature was evidently used by the {\sc xspec} fitting routine 
to describe the `valley' in the energy range 6-9~keV where the soft and hard spectral components intersect one another. We thus do not 
discuss this model further. For completeness, instead, we tried to fit the two component model used in the present 
paper to the combined \swift\,/XRT and \nustar\ data reported previously by \citet{bhalerao15}, where the cyclotron line was discovered. 
We used for the fit the same spectra extracted by these authors, as the latters are also  among the 
collaborators of the present publication.

A  fully reasonable fit was obtained with $\chi^2_{\rm red}$/d.o.f.=0.85/144, $N_{\rm H}$=1.0$\pm$0.3~cm$^{-2}$, 
$kT_{\rm BB}$=1.07$\pm$0.04~keV, $E_{\rm cut}$=6.7$^{+0.9}_{-6.7}$~keV, and $E_{\rm fold}$=4.6$\pm$0.1~keV (we had to fix $\Gamma$=-2.5 in the 
fit as this parameter could not be constrained and the fit favored a largely negative power-law photon index). 
The presence of the cyclotron line remained clearly evident even from the residuals obtained 
with this model and we measured a centroid energy of 16.9$\pm$0.5~keV, a depth of 0.5$\pm$0.1~keV, and a 0.5-50~keV X-ray flux of 
3.5$\times$10$^{-11}$~\eflux in agreement with previous findings 
\citep{bhalerao15}. We thus concluded that, although we could not confirm 
the presence of the cyclotron line in \src\ with the newly obtained, strictly simultaneous \xmm\ and \nustar\ observations, this feature did not 
disappear from the previous data-set when a different spectral model was used for the fit (note that the average flux of the 
source in the past and present observations is virtually the same).

\section{Spectral analysis with a physical model}
\label{sec:physical}

As SFXTs are believed to host young, highly magnetized accreting NSs and the spectral analysis above showed the presence of both a thermal and 
a non-thermal component in the spectrum of \src,\ we also attempted a description of the X-ray emission from this object with a more physical 
model. In particular, we adopted the bulk and thermal Comptonization model (BW) described in detail by \citet{ferrigno09}. 
The model is based on the 
original calculation of \citet{becker07}, who computed the X-ray emission emerging from a cylindrical accretion column, typical 
of NS hosted in young high-mass X-ray binaries (and thus also in SgXBs). The model computes the bulk and thermal Comptonization of 
the seed photons that are produced by the bremsstrahlung, cyclotron, and blackbody processes taking place within an accreting plasma, 
characterized by a constant magnetic field and electron density. In this situation, the distributions in energy of the cyclotron 
and bremsstrahlung photons are generally different. However, in those cases where the cyclotron energy is comparable to the temperature 
of the plasma, the cyclotron emission becomes the preferred cooling channel. The blackbody emission is assumed to be concentrated 
toward the bottom of the accretion column. We refer the reader to \citet{ferrigno09} and \citet{becker07} 
for further details.
\begin{table*}[htb!]
{\fontsize{6.3}{4}\selectfont 
\centering
\caption{Results of the fits to the combined \xmm\ and \nustar\ spectra extracted in the time resolved intervals with the phenomenological 
model. In each case, the normalization constant with no indicated uncertainties is the one of the instrument used as a reference in the fit.}
\label{tab:fits}
\bgroup
\def\arraystretch{2.5}
\begin{tabular}{@{}llllllllllllll@{}}
\hline
Parameter & All$^{a}$ & Quiescent & 1 & 2 & 3 & 4 & 5 & 6 & 7 & 8 & 9 & 10$^{b}$ & 20     \\
\hline                  
$N_{\rm H}$ &0.8$\pm$0.06 & 1.14$\pm$0.06 & 1.8$\pm$0.3 & 0.96$\pm$0.15 & 1.8$\pm$0.3 & 1.1$\pm$0.4 & 1.6$\pm$0.3 & 1.8$\pm$0.2 & 1.7$\pm$0.2 & 1.2$\pm$0.5 & 1.3$_{-0.4}^{+0.6}$ & 1.4$\pm$0.3 & 1.2$\pm$0.2  \\
$kT_{\rm BB}$ &1.11$_{-0.01}^{+0.03}$  & 1.08$\pm$0.03 & 0.53$\pm$0.01 & 1.37$\pm$0.08 & 1.3$\pm$0.1 & 1.35$\pm$0.14 & 1.7$_{-0.3}^{+0.5}$ & 1.6$\pm$0.2 & 1.4$_{-0.1}^{+0.2}$ & 1.1$\pm$0.2 & 1.1$_{-0.1}^{+0.2}$ & 1.04$\pm$0.1 & 0.90$\pm$0.07   \\
$N_{\rm BB}$ &0.41$\pm$0.03 & 0.12$\pm$0.01 & 12.6$_{-6.7}^{+18.1}$ & 1.1$\pm$0.2 & 4.8$_{-0.8}^{+1.0}$ & 9.9$_{-5.0}^{+4.0}$ & 6.8$_{-4.2}^{+8.4}$ & 4.8$_{-2.6}^{+3.8}$ & 4.2$_{-2.4}^{+2.9}$ & 12.0$\pm$5.5 & 11.1$\pm$4.7 & 0.7$\pm$0.4 & 1.4$\pm$0.3 \\
$\Gamma$ &0.42$\pm$0.09  & -1.3$_{-0.8}^{+0.5}$ & 0.7$\pm$0.1 & -1.7$_{-0.7}^{+0.6}$ & -0.13$_{-0.4}^{+0.3}$ & 0.3$_{-1.2}^{+0.4}$ & 0.8$_{-0.2}^{+0.1}$ & 0.92$_{-0.10}^{+0.09}$ & 0.8$\pm$0.1 & 0.6$_{-0.5}^{+0.3}$ & 0.5$_{-1.5}^{+0.5}$ & 0.7$_{-0.4}^{+0.3}$ &  0$_{-0.5}^{+0.4}$  \\
$E_{\rm cut}$ & 15.2$\pm$0.03  & 12.2$_{-0.9}^{+0.8}$ & 17.3$\pm$1.1 & 13.4$\pm$0.6 & 16.1$\pm$0.7 & 14.6$_{-1.5}^{+1.2}$ & 16.3$_{-1.3}^{+0.8}$ & 16.0$\pm$0.7 & 15.7$\pm$0.8 & 15.5$_{-1.2}^{+1.4}$ & 15.2$_{-3.9}^{+4.1}$ & 16.8$_{-3.5}^{+5.9}$ & 14.3$_{-1.3}^{+1.5}$  \\
$E_{\rm fold}$ & 6.8$\pm$0.03  & 4.6$_{-0.8}^{+1.0}$ & 7.2$\pm$1.3 & 3.9$_{-0.5}^{+0.7}$ & 4.7$\pm$0.6 & 6.2$_{-0.8}^{+1.6}$ & 5.8$_{-0.7}^{+0.9}$ & 6.9$\pm$0.8 & 7.0$\pm$0.8 & 6.3$_{-1.4}^{+2.0}$ & 8.5$\pm$4.6 & 6.0$_{-6.0}^{+6.2}$ & 7.1$_{-1.6}^{+2.5}$  \\
$F_{\rm 0.5-10~keV}$ &1.61E-11 & 1.8E-12 & 6.3E-11 & 5.0E-11 & 2.2E-10 & 7.9E-10 & 2.3E-9 & 1.6E-9 & 7.6E-10 & 5.6E-10 & 3.8E-10 & 3.1E-11 & 1.7E-11  \\
$F_{\rm 10-30~keV}$ & 1.98E-11  & 2E-12 & 1.3E-10 & 1.3E-10 & 4.6E-10 & 1.3E-9 & 3.7E-9 & 2.5E-9 & 1.3E-9 & 9.9E-10 & 6.4E-10 & 8.3E-11 & 3.1E-11  \\
$\chi^2_{\rm red}$/d.o.f. & 1.11/986  & 1.09/479 & 1.27/414 &  1.07/331 &0.94/338 & 1.04/248 & 0.91/368 & 1.04/607 & 0.91/536 & 1.04/208 & 0.90/185 & 0.86/210 & 1.06/258  \\
$C_{\rm pn}$ & ---  & 1.0 & 1.00$\pm$0.05 &  1.03$\pm$0.06 &0.97$\pm$0.05 & 0.95$\pm$0.07 & 0.98$\pm$0.06 & 0.97$\pm$0.04 & 1.03$\pm$0.04 & 1.02$\pm$0.08 & 0.98$\pm$0.07 & 0.94$\pm$0.05 & 1.00$\pm$0.05  \\
$C_{\rm MOS1}$ & ---  & 1.12$\pm$0.03 & 0.96$\pm$0.07 & 1.01$\pm$0.08 &1.03$\pm$0.10 & 0.85$\pm$0.17 & --- & --- & --- & --- & --- & 0.90$\pm$0.09 & 1.0$^{c}$  \\
$C_{\rm MOS2}$ & 1.0  & 1.34$\pm$0.03 & 1.0 & 1.0 &1.0 & 1.0 & 1.0 & 1.0 & 1.0 & 1.0 & 1.0 & 1.0 & 1.0  \\
$C_{\rm RGS1}$ & 1.21$\pm$0.07  & --- &--- & --- & --- & --- & --- & --- &--- & --- & --- & --- & ---  \\
$C_{\rm RGS2}$ & 1.10$\pm$0.10  & --- & --- & --- & --- & --- & --- & --- &--- & --- & --- & --- & ---  \\
$C_{\rm FPMA}$ & 0.75$\pm$0.01  & 0.93$\pm$0.04 & 0.99$\pm$0.07 & 1.06$\pm$0.09 &1.08$\pm$0.07 & 1.01$\pm$0.07 & 1.08$\pm$0.06 & 1.11$\pm$0.04 & 1.08$\pm$0.05 & 1.14$\pm$0.10 & 0.95$\pm$0.08 & 1.6$\pm$0.2 & 1.00$\pm$0.09 \\
$C_{\rm FPMB}$ & 0.79$\pm$0.02  &1.11$\pm$0.09  & 1.05$\pm$0.07 & 1.03$\pm$0.06 & 1.14$\pm$0.07 & 1.00$\pm$0.07 & 1.13$\pm$0.06 & 1.11$\pm$0.04  &1.11$\pm$0.5 & 1.14$\pm$0.09 & 1.00$\pm$0.09 & 1.8$\pm$0.2 & 1.01$\pm$0.10 \\
\hline \\
\end{tabular}
\egroup
\begin{itemize} 
\scriptsize
\item[$^{\mathrm{a}}$:] For these spectra a thin iron line was also included in the fit. 
The centroid energy of the line is 6.39$\pm$0.03~keV and the corresponding equivalent width is 0.073$\pm$0.016~keV. 
\item[$^{\mathrm{b}}$:] For these spectra we also included in the fit an absorption line at 7.21$\pm$0.14~keV with an equivalent width of 
0.23$_{-0.06}^{+0.11}$~keV and a width of 0.2$\pm$0.1~keV. 
\item[$^{\mathrm{c}}$:] This value was fixed in the fit as the fit was insensitive to any variation of 
this parameter within reasonable boundaries. 
\end{itemize} 
}
\end{table*}

The BW model has six free parameters\footnote{In all cases, we fixed the NS mass and radius to the canonical values, 
$M_{\rm NS}$=1.4~$M_{\odot}$ and $R_{\rm NS}$=10$^6$~cm. These parameters could not be constrained in any of the fits.}: 
the mass accretion rate $\dot{M}$, the radius of the accretion 
column $r_0$, the temperature of the electrons $T_{\rm e}$, the magnetic field strength $B$, the photon diffusion 
parameter $\xi$, and the Comptonization parameter $\delta$. The last two parameters are defined as: 
\begin{eqnarray}
& & \xi = \frac{\pi r_0 m_{\rm p}c}{\dot{M}(\sigma_{\vert\vert}\sigma_{\perp})^{1/2}}      , \\
& & \delta = 4 \frac{y_{\rm bulk}}{y_{\rm thermal}}, 
\end{eqnarray} 
where $m_{\rm p}$ is the proton mass, $c$ is the speed of light, 
$\sigma_{\vert\vert}$ , ($\sigma_{\perp}$) is the electron scattering cross section for photons 
propagating parallel (perpendicular) to the magnetic field direction, and 
$y_{\rm bulk}$ ($y_{\rm thermal}$) describes the average fractional energy change 
experienced by a photon before it escapes through the walls of the accretion column 
due to the bulk (thermal) Comptonization. As in this model the normalization is regulated 
by both $\dot{M}$ and $r_0$ and in all fits we fixed the value of 
the mass accretion rate to the one estimated from the source broad-band X-ray luminosity 
(0.5-30~keV) with the usual formula $L_{\rm X}$=$GM_{\rm NS}\dot{M}$/$R_{\rm NS}$ and left 
$r_0$ free to vary. 

Due to all physical assumptions in the treatment presented by \citet{becker07},  
the BW model is only suited to describe the X-ray energy distribution of X-ray pulsars with a luminosity 
larger than a certain critical value. The latter corresponds to the luminosity at which 
the radiation emerging from the NS surface is able to
stop the accreting matter infalling through the accretion column via a radiation-dominated shock. 
The precise value of this critical luminosity is highly debated, but the most recent calculations 
presented by \citet{mushtukov15} show that it should not be lower than $\sim$2$\times$10$^{36}$~erg~s$^{-1}$ 
for a NS endowed with a magnetic field similar to that suspected for \src\ (i.e. $\sim$1.5$\times$10$^{12}$~G, 
see Sect.~\ref{sec:intro}). At the distance of \src,\ the above luminosity corresponds to an X-ray flux of 
$\sim$10$^{-9}$~\eflux. Based on the results obtained in Sect.~\ref{sec:pheno} and on the considerations above, 
we thus performed fits to spectra 3-9 in Table~\ref{tab:fits} with the BW model only. 
A blackbody was included in all fits to take into account the presence of the thermal 
component emitted from the BW model. The results of these fits are summarized in Table~\ref{tab:physical}.  
We note that the measured parameters of the BW model are comparable to the values expected 
for a highly magnetized NS ($\sim$10$^{12}$~G). The estimated radius of the accretion column is substantially 
lower than that measured in the case of the brightest X-ray pulsars \citep[up to several hundreds of meters; see, e.g.,]
[and references therein]{walter2015}, but in line with the prediction for dimmer systems, such as Her X-1 \citep{becker07}. 

On the right side of Fig.~\ref{fig:otherspectra} we show a comparison between 
the fits with the physical and phenomenological models to the spectra 6 and 9, as these were previously considered as two representative cases   
for the spectral changes occurring in the source X-ray emission during the decay from the second flare (when the most 
prominent spectral variability was recorded). 
\begin{table*}
\tiny
\label{tab:physical}
\centering
\caption{Results of the fits to the combined \xmm\ and \nustar\ spectra with the physical model introduced 
in Sect.~\ref{sec:physical}.} 
\begin{tabular}{llllllll}
\hline
Parameter & 3 & 4 & 5 & 6 & 7 & 8 & 9 \\
\hline                  
$N_{\rm H}$ (1E22~cm$^{-2}$) &1.9$_{-0.2}^{+0.3}$ & 1.3$_{-0.2}^{+0.3}$ & 1.2$_{-0.3}^{+0.2}$ & 1.2$_{-0.3}^{+0.2}$ & 1.1$\pm$0.1 & 1.0$\pm$0.4 & 1.0$\pm$0.3 \\
$kT_{\rm BB}$ (keV) & 1.5$\pm$0.1  & 1.7$\pm$0.2 & 1.6$\pm$0.3 & 1.5$\pm$0.2 & 1.4$\pm$0.1 & 1.4$_{-0.2}^{+0.4}$ & 1.2$\pm$0.2 \\
$N_{\rm BB}$ & 3.9$_{-0.5}^{+0.7}$ & 8.3$_{-2.1}^{+3.1}$ & 25$_{-10}^{+18}$ & 19.3$_{-6.3}^{+9.2}$ &  13.6$_{-3.2}^{+6.0}$ & 8.9$_{-5.3}^{+8.3}$ & 11.4$_{-3.6}^{+6.1}$\\
$\xi$ & 5.0$_{-0.7}^{+8.0}$  & $<$15 & 4.5$_{-3.0}^{+4.2}$ & 5.4$_{-2.5}^{+3.6}$ & 5.7$_{-5.0}^{+6.8}$ &  5.7$_{-3.6}^{+6.6}$ & $<$5.2 \\
$\delta$ & 0.8$_{-0.4}^{+0.2}$  & 0.4$_{-0.2}^{+0.5}$ & 0.7$_{-0.3}^{+0.9}$ & 0.6$_{-0.3}^{+0.5}$  & 0.3$_{-0.1}^{+0.4}$ & 0.7$_{-0.5}^{+0.8}$ & $<$1.0 \\
$\dot{M}_{1E17 g/s}^{a}$ & 0.05 & 0.17 & 0.48 & 0.32 & 0.16 & 0.12 & 0.08 \\
$T_{\rm e}$ (keV) & 3.2$_{-0.2}^{+0.3}$  & 3.4$\pm$0.3 & 2.9$_{-0.4}^{+0.3}$ & 3.0$\pm$0.3 & 3.4$\pm$0.3 & 2.8$_{-0.4}^{+0.6}$ & 3.7$_{-0.7}^{+0.9}$\\
$r_0$ (m) & $<$20 & 47$_{-25}^{+14}$ & 65$\pm$33 & 53$_{-18}^{+21}$ & 35$_{-23}^{+18}$ & 18$_{-18}^{+22}$ & 19$_{-19}^{+28}$ \\
$F_{\rm 0.5-10~keV}^{b}$ & 2.2E-10 & 7.8E-10 & 2.4E-9 & 1.6E-9 & 7.6E-10 & 5.6E-10 & 3.8E-10 \\
$F_{\rm 10-30~keV}^{b}$ & 4.6E-10 & 1.3E-9 & 3.6E-9 & 2.5E-9 & 1.3E-9 & 9.7E-10 & 6.6E-10 \\
$\chi^2_{\rm red}$/d.o.f. & 0.94/338  & 1.03/248 & 0.92/375 & 1.08/605 & 0.93/534 & 1.05/207 & 0.90/185 \\
$C_{\rm pn}$ & 0.96$\pm$0.05 & 0.96$\pm$0.07 & 0.95$\pm$0.05 & 0.97$\pm$0.04 & 1.04$\pm$0.04 & 1.02$\pm$0.07 & 0.98$\pm$0.07 \\
$C_{\rm MOS1}$ & 1.0$\pm$0.1 & 1.0$\pm$0.1 & --- & --- & --- & --- & --- \\
$C_{\rm MOS2}^{c}$ & 1.0 & 1.0 & 1.0 & 1.0 & 1.0 & 1.0 & 1.0 \\
$C_{\rm FPMA}$ & 1.08$\pm$0.08  & 1.02$\pm$0.07 & 1.06$\pm$0.06 & 1.11$\pm$0.04 & 1.07$\pm$0.04 & 1.14$\pm$0.09 & 0.95$\pm$0.09 \\
$C_{\rm FPMB}$ & 1.13$\pm$0.07  &1.02$\pm$0.07  & 1.11$\pm$0.06 & 1.11$\pm$0.04 & 1.11$\pm$0.05 & 1.13$\pm$0.09 & 1.00$\pm$0.08 \\
\hline \\
\end{tabular}
\begin{itemize} 
\scriptsize
\item[$^{\mathrm{a}}$:] The mass accretion rate is derived from the 0.5-30~keV X-ray 
flux using standard equations for the NS accretion, that is, 
$L_{\rm X}=GM_{\rm NS}\dot{M}/R_{\rm NS}$ and $L_{\rm X}=4\pi F_{\rm 0.5-30~keV} d^2$ 
(here $M_{\rm NS}$ and $R_{\rm NS}$ are the mass and radius of the NS, whilst $d$ is the source 
distance). 
\item[$^{\mathrm{b}}$:] The flux is given in \eflux.\ 
\item[$^{\mathrm{c}}$:] The MOS2 was used in all case as the reference instrument, therefore the 
corresponding normalization constant has been fixed to unity. 
\end{itemize} 
\end{table*}

\section{Optical and infrared observations}
\label{sec:others} 

We also obtained optical and infrared data during the long observations performed with 
\xmm\ and \nustar.\ Data were collected with: 
\begin{itemize}
\item the Berlin Exoplanet Search Telescope (BEST\,II), which is a 25\,cm aperture Baker-Ritchey-Chr\'{e}tien system, 
using a KAF\,16801 $4096\times4096$ pixel CCD with a pixel size of 9\,$\mu$m and a field of view 
of $1.7^{\circ}\times1.7^{\circ}$ \citep[see,][for all relevant information]{kabath09}.
\item the Bochum Monitoring Telescope\footnote{http://www.astro.rub.de/Astrophysik/BMT\_en.html} 
(BMT), which is a 40\,cm Coud\'{e} telescope, featuring a SBIG STL-6303 
CCD with $3072\times2048$ pixel (each sized 9\,$\mu$m) with a field of view of 
$41.2' \times 27.5'$ \citep{ramolla13}. 
\item the Infrared Imaging System (IRIS) telescope, which is a 80\,cm Nasmyth telescope 
equipped with a HAWAII-1 detector. 
The telescope field of view is $12.5' \times 12.5'$ (with a pixel size of $0\farcs74 \times 0\farcs74$), 
and a filter wheel equipped with standard 2MASS $J,H,K$ and narrow band filters \citep{hodapp10}.
\end{itemize}

\src\ was observed with all mentioned instruments on 2015 March 21 in Johnson BVRI 
filters and 2MASS Jn and Ks filters. The coverage in the different filters is specified in Table~\ref{table:logOIR}. 
\begin{table}
\caption{Log of optical and NIR observations of \src\ \label{table:logOIR}}
\begin{center}
\begin{tabular}{lllll} 
\hline
Filter & Telescope & Start time & End time & Exptime  \\
       &           & (UT)       & (UT)     & (s)      \\
B-V    & BMT       & 8:25:58    & 08:57:16 & 1900     \\
R-I    & BEST      & 08:35:51   & 09:18:12 & 2600     \\
Ks-Jn  & IRIS      & 09:06:58   & 10:06:52 & 3600     \\
\hline
\end{tabular}
\end{center}
\end{table}
All photometric data were reduced using standard Image Reduction and Analysis Facility (IRAF) bias, dark, 
and flat field correction routines. Astrometry was provided by {\sc scamp} \citep{bertin06} and before combining a set 
of multiple exposures into a final frame, the data were re-sampled onto a common WCS frame using {\sc swarp} 
\citep{bertin02}. The photometry was performed on combined frames using a $7\farcs5$ aperture.
To obtain absolute calibration in the optical bands, the Landolt star fields SA\,104 and SA\,107 
\citep{landolt09} were observed. The field star fluxes were cross-calibrated with the 
Landolt photometry, taking into account the airmass dependent extinction based on the atmospheric profile of 
the Cerro Paranal site obtained by \citet{patat11}. In the infrared, available 2MASS AAA photometry of 
field stars in our frames was used to directly cross calibrate the photometry.

We report all photometric results of the averaged frames with min/max rejection in Table~\ref{tab:opt}. 
The magnitude of the source in the observations with different filters was estimated using isolated field 
stars and the UCAC-4 photometry. In the Table we indicate the exposure time of single frame observations for 
each filter and the number of frames combined in each case to obtain the final results. 
For comparison we also report the results on all filters that we could find in the literature 
for \src.\  
\begin{table}
\scriptsize
\caption{Results obtained from the optical and NIR photometry of \src.\ 
\label{tab:opt}}
\begin{center}
\begin{tabular}{@{}lllll@{}} 
\hline
Filter & NFrames & Exp. time(s) & \multicolumn{2}{c}{Magnitude} \\
       &         &            & (this work) &  (literature) \\
B      & 11      & 60         & 14.62$\pm$0.05       & 14.44$\pm$0.05$^a$ \\
V      &  7      & 60         & 12.89$\pm$0.05       & 12.65$\pm$0.05$^a$ \\
R      &  9      & 60         & 11.76$\pm$0.05       & $<$11.9$^a$ \\
I      &  9      & 60         & 10.39$\pm$0.05       &           \\
Jn     & 20      & 10         &  8.77$\pm$0.05       &  8.71$\pm$0.02$^a$; 8.791$\pm$0.021$^b$ \\
Ks     & 20      & 10         &  8.21$\pm$0.05       &  7.99$\pm$0.02$^a$; 8.018$\pm$0.026$^b$ \\
\hline
\end{tabular}
\begin{itemize} 
\scriptsize
\item[$^{\mathrm{a}}$:] Data from \citet{pellizza06}. 
\item[$^{\mathrm{b}}$:] Data from the 2MASS catalogue \citep{2mass}. 
\end{itemize} 
\end{center}
\end{table}

As all the optical and IR observations of \src\ were obtained several hours before the bright outburst from the 
source detected by \xmm\ and \nustar,\ the results reported in Table~\ref{tab:opt} describe 
the properties of the supergiant companion when the system was in a quiescent state. 
Comparison with previous results in the literature reveals that the supergiant 
was globally fainter during our observations. However, the differences with the previously reported 
magnitudes are very limited and likely consistent with the micro-variability observed in other supergiant systems 
\citep[see, e.g, the case of IGR\,J16465-4507;][]{chaty16}.

\section{Discussion}
\label{sec:discussion}

In this paper we report on a long multi-wavelength campaign observation of the SFXT prototype 
\src.\ This source is known to display one of the most extreme levels of X-ray variability among 
other objects in the same class. A remarkable dynamic range in the 
X-ray luminosity was also observed. 

\src\ was initially caught by both \xmm\ and \nustar\ during an 
extended quiescent period, which covered the first $\sim$120~ks of the observations and was characterized 
by a luminosity of $\sim$6$\times$10$^{33}$~\ergsec. During this period the statistics of the data were too 
low to carry out a time-resolved spectral analysis, and thus a single spectrum was extracted for all \xmm\ 
instruments and for the two FMs of \nustar.\ The source spectral energy distribution in the X-ray domain 
could be well-described by using a combination of a thermal component, most likely associated with the  
emission from the neutron star surface, and a non-thermal component extending up to $\sim$40~keV. The latter is 
usually ascribed in similar systems to Comptonization processes occurring within the accretion column of 
the neutron star. The peculiarly high emission around 10-20~keV made it very difficult to fit the non-thermal 
component with any other phenomenological model other than the {\sc highecut*pow} in {\sc xspec}. This model provided 
reasonably good fit to all spectra extracted from both the \xmm\ and \nustar\ data. 

A bright outburst abruptly interrupted the source quiescent state towards the end of the X-ray observations and 
lasted for approximately 7~ks. The event comprised three distinct fast flares, among which the first was the faintest 
and the other two achieved a luminosity $\gtrsim$1600 times brighter than quiescence. Significant spectral 
variability was observed, especially during the second flare. In particular, the broad-band fits realized by combining the 
time-resolved \xmm\ and \nustar\ data showed that there was a modest (but significant) increase of the absorption column 
density during the rise to the peak of the flare (a factor of $\sim$1.5), followed by a drop of the column density 
immediately after the source reached the peak X-ray flux. This is reminiscent of what was observed during the bright 
flare from the SFXT IGR\,J18410-0535 \citep{bozzo11}. On that occasion the increase in the absorption column density during the 
onset of the flare was much larger (by a factor of $\sim$50) however it was also observed to drop significantly at the peak of the event. 
The interpretation in this case was that a large clump in the wind of the supergiant companion encountered the neutron star and 
partially obscured the X-ray source before being accreted onto the compact object. The increase in the absorption column 
density and the total X-ray luminosity released during the flare were used to estimate the physical projected size of the clump and its mass. 
The drop in the absorption column density around the peak of the flare could be explained by assuming that the clump material became  
significantly ionized when the incident X-ray radiation exceeded a certain threshold. 

The above interpretation could also be applied to explain the 
event observed from \src.\ However, why a much lower increase in the column density was measured   
in the present case compared to the flare displayed by IGR\,J18410-0535 remains puzzling, although the peak flux in the latter event was approximately
a factor of ten lower than that recorded during the outburst of \src.\ A viable solution could be to assume that the flare from \src\ was observed from 
an unfavorable geometry in which the clump approaching the neutron star was not located along the line of 
sight to the observer and thus gave rise to a limited increase in the measured absorption column density. 
The presence of different flares within the same outburst could be ascribed to the presence of structured clumps or to the accretion 
of multiple smaller clumps impacting one after the other onto the compact object \citep{walter07}. 

This scenario would be in line with our current understanding of the SFXT phenomenology 
(see Sect.~\ref{sec:intro}), however the current consensus is that clumps alone cannot be the sole explanation for all the 
peculiar properties of the X-ray behavior of these sources \citep[see, e.g.,][]{bozzo15}. It remains particularly difficult to explain why \src,\ as well as other SFXTs, displays on average 
a much lower luminosity compared to the value that would be expected if they were classical SgXBs. The monitoring we performed with 
XRT during the entire orbital revolution of \src,\ during which also the \xmm\ and \nustar\ observations were carried out, showed that the 
source remained in a relatively faint X-ray state for most of the time. The average flux of 2.7$\times$10$^{-12}$~\eflux recorded by XRT 
corresponds to an X-ray luminosity of 4$\times$10$^{33}$~\ergsec at the distance of \src.\ As pointed out by \citet{romano14} and 
\citet{bozzo15}, this is orders of magnitudes lower than the luminosity expected from a classical wind-fed SgXBs having similar 
orbital parameters to those of \src\ \citep[see also][]{lutovinov13}. It should thus be argued that some mechanism is at work in the SFXTs 
inhibiting accretion for most of the time. Among the different possibilities proposed so far and summarized in Sect.~\ref{sec:intro}, 
none seem able to satisfactorily explain all observational properties shown by all SFXTs so far. As discussed in \citet{walter2015}, it is 
unlikely that the settling accretion regime alone could produce a dynamic X-ray range as high as the one displayed by \src,\ as it would 
require a large systematic difference in the stellar winds of the SFXT supergiant companions with respect to those in 
classical systems which seem not to be supported by the available observations. The gating models typically require a large neutron star 
magnetic field to be able to reproduce an SFXT-like behavior. As it was recently shown by \citet{bozzo16}, it is particularly challenging 
to explain within the assumptions of this model the onset of very sporadic bright outbursts if the neutron star magnetic field is not 
$\gtrsim$10$^{13}$-10$^{14}$~G. Although in the present paper we could not confirm the presence of a cyclotron line in the X-ray spectrum 
of \src\ as found previously by \citet{bhalerao15}, the detection of this feature is not affected by the specific model adopted to 
fit the source broad-band spectrum, and thus the estimate of the source's relatively low magnetic field ($B\simeq1.5\times10^{12}$~G) 
seems robust. At present, we thus conclude that the X-ray behavior displayed by \src\ (as well as by the other extreme SFXTs) 
is challenging all presently proposed theoretical models\footnote{We note that the suggestion by \citet{gimenez16} according to which 
\src\ could spend most of its time in the so-called supersonic propeller regime critically depends on the value of the source 
spin period, a parameter that is not known yet (two values were suggested so far and never confirmed, see Sect.~\ref{sec:intro}).} 
and a satisfactorily explanation for the exceptional X-ray variability 
of this object requires additional progress in the development of current models. 

Assuming that the neutron star hosted in \src\ is similar to the X-ray pulsars identified in many other SgXBs and high mass X-ray binaries (HMXBs) in general, we also 
attempted to model its high energy emission in Sect.~\ref{sec:physical} with the physical model proposed by \citet{becker07} plus the contribution 
of the thermal emission coming from the neutron star surface. In the BW model the higher energy photons are  
produced by the free-free interactions in the magnetic field of the pulsar and then up-scattered within the accretion column. 
With a neutron star magnetic field of $\sim$10$^{12}$~G, electrons in the accretion stream 
populate the first Landau level by collisional excitation, largely affecting the X-ray spectrum emerging from 
the column. Calculations by \citet{riffert88} showed that in the optically thin regime an enhanced emission is 
produced at energies close to the local cyclotron energy, while the thermal 
breemstrahlung dominates the emission at much lower energies. In the simplified approach of \citet{becker07}, the 
breemstrahlung and the cyclotron emission are considered as separate contributions in order to linearize an otherwise 
very complicated process. For the input spectrum these authors assume  a thermal breemstrahlung as well as a delta-shaped emission for the 
cyclotron component. The Comptonization of both components within the accretion column is then realized through a Green's function 
that takes into account the thermal and bulk Comptonization. The implementation of the BW 
model\footnote{http://www.isdc.unige.ch/$\sim$ferrigno/images/Documents/ BW\_distribution/BW\_cookbook.html.} 
within {\sc xspec} was discussed by \citet{ferrigno09} and first tested on the HMXB 4U~0115+63. 
The pronounced emission detected from \src\ around 10-20~keV, which was highlighted in Sect.~\ref{sec:pheno} as a 
peculiar feature of the source spectral energy distribution, is elegantly reproduced in the BW model by the broadened 
cyclotron emission. This corresponds to the large peak visible in the right-hand plots of Fig.~\ref{fig:otherspectra}  
that covers approximately the energy range 10-20~keV. Note that, as should be expected in this model, the broadened cyclotron emission is 
centered around the centroid energy of the cyclotron feature previously detected in the X-ray spectrum 
of \src\ \citep{bhalerao15}. 

It should be remarked that the Comptonization of the black body seed photons coming from the base of the 
accretion column and self-consistently accounted for in the BW model yields a negligible contribution to the computed 
X-ray emission. As discussed in \citet{ferrigno09}, the presence of an evident, additional soft thermal component 
in the spectrum of the source can be interpreted assuming the presence of an extended halo on the NS surface. This could be produced 
by either material arriving on regions of the neutron star surface external to the column during the accretion process or 
photons in the accretion stream that heat a sufficiently large fraction of the neutron star surface. From the results of the 
time resolved spectral analysis carried out with the BW model (see Table~\ref{tab:physical}), we can conclude that the 
relative intensity of the black-body and cyclotron emission components varies with the source luminosity, with the black-body 
being more prominent when the source is dimmer. This is compatible with the behavior of thermal and non-thermal components 
observed in other X-ray pulsars \citep[see][and references therein]{ferrigno09} and with our interpretation that the black-body 
emission is due to the continuous heating of the neutron star surface. We mentioned in Sect.~\ref{sec:physical} that the 
spectral parameters measured from the fits of the physical model to the combined \xmm\ and \nustar\ data lie between the 
usual boundaries determined for a number of other highly magnetized X-ray pulsars \citep[see, e.g.,][and references therein]{walter2015}  
and therefore cannot easily help in understanding why SFXTs behave in such an atypical way compared to other wind-fed HMXBs. 
As discussed and analyzed by \citet{shakura13}, the pulse profile of these sources, together with their energy and time dependence, 
are probes of the different accretion mechanisms and geometries. This could help to discriminate among the different theoretical 
models proposed to interpret the SFXT behavior, however these investigations are at present hampered by the lack of spin period 
measurements for most of the SFXTs (see Sect.~\ref{sec:intro}). Longer and deeper X-ray observations are thus needed in order to eventually 
confirm the tentative spin period detections reported for \src\ and to sensitively search for pulsations in all other SFXTs. 
Note that the presence of a hot, relatively confined black-body component on the surface of the NS hosted in \src,\ 
as well as in a number of other SFXTs, would suggest that pulsations can be expected from these sources and might have gone 
undetected so far mostly due to the very long spin period \citep[see, e.g.,][and references therein]{bozzo08,bozzo10,sidoli09c}. 
A similar conclusion was reached by \citet{walter16}, who proposed that the spacing among the different flares usually detected
within the SFXT structured outbursts (generally a few ks) could be an indication of the NS spin period. If the magnetic 
and rotation axes of the NS are closely aligned and the radiation is beamed in unfavourable directions, then we might 
only be able to see pulsations during the brightest outbursts, while for most of the time the source remains barely detectable.   

Interestingly, if we use the most recent ephemeris available for \src\ \citep[orbital period 4.92693$\pm$0.00036~d and 
periastron passage at 53732.65$\pm$0.23~MJD;][]{smith14}, it turns out that the bright outburst observed by \xmm\ and \nustar\ 
falls at the expected phase of the pariastron passage. This confirms previous findings that most of the 
outbursts displayed by \src\ occur when the neutron star is closer to the supergiant companion 
and that the orbit of this system could be characterized by a non-negligible eccentricity  
\citep[see, e.g.,][and references therein]{drave14,romano15}. Although this eccentricity could help in enhancing the X-ray dynamic range 
achievable by \src,\ it cannot be the only explanation for the extreme behavior displayed by this source, as the system 
orbital period is relatively short and only a limited eccentricity of $\lesssim$0.2-0.3 can be expected \citep{walter2015,gimenez16}.  

In SFXTs with orbital periods of only a few days and characterized by a non-negligible 
eccentricity, as for example \src\ and IGR\,J16479-4514,  
it was suggested that the neutron star could get close enough to the supergiant to largely slow down its wind 
through X-ray photoionization and lead to the formation of temporary accretion disks 
\citep[see][and references therein]{ducci10}. In the case of \src,\ the presence of such structures was first 
suggested by \citet{romano15b} based on the peak luminosity achieved during an outburst on 2014 October 10 
that was too high to be produced within a wind-fed system. No direct evidence of disks in SFXTs has been 
reported so far. In the \xmm\ observations of \src\  during the time interval 10 (see Fig.~\ref{fig:hr}) we found an 
intriguing feature at 7.2~keV. This is reminiscent of the iron absorption lines usually observed in high inclination 
low mass X-ray binaries when material is pulled out from the disk and the X-ray radiation passes through it,  
ionizing this absorber before arriving at the observer \citep[see, e.g.,][and references therein]{trigo13}. 
On one hand, the presence of this feature could indicate the presence of at least a temporary disk-like 
structure\footnote{In principle, the iron emission 
line detected when the \xmm\ and \nustar\ data are averaged over the entire observational period could also be formed 
in an accretion disk. However, the fact that the line is thin and compatible with what is usually observed from other 
SFXTs and wind-fed HMXBs suggests that it is more likely produced due to fluorescence in the supergiant wind material surrounding 
the compact object and illuminated by its X-ray emission \citep[see, e.g.,][]{walter2015}.} 
around the neutron star in \src,\ opening up the possibility that a different accretion mechanism, poorly explored so far, 
could play a role in regulating the SFXT X-ray variability. On the other hand, it is quite unlikely that \src\ is observed at 
high inclination angles, as the system is characterized by a relatively small orbital separation but no eclipses are 
detected in its long term lightcurve \citep{romano14b}. An alternative possibility is that the material 
filtering the X-ray radiation could be associated with the clump being accreted and ionized during the first two flares 
of the source X-ray outburst. While this interpretation is much more in line with the scenario depicted above to interpret the 
overall behavior displayed by \src,\ we remark that (to the best of our knowledge) no similar features have so far been observed 
in other wind-fed HMXBs. The limited statistics of the \xmm\ data during the relatively short and faint time interval 10, 
combined with the lack of complete coverage of this interval by \nustar,\ have prevented us from performing a more refined 
study of this feature.  

Finally, we also reported on the results of optical and IR observations carried out during the same orbital revolution of \src\ 
observed in X-rays with \swift\,/XRT, \xmm,\ and \nustar.\ These additional observations captured the system during the 
quiescent period and did not reveal any peculiar changes in the properties of the supergiant companion that could provide help in 
investigating the mechanism triggering the bright outburst observed in X-rays. The measured limited changes in the optical and 
IR magnitudes of the source are compatible with the expected micro-variability of supergiant stars. A similar consideration 
applies to the UVOT data in the UV energy range, which provided measurements of the magnitudes in different filters compatible 
with previous results reported in the literature. We note that the results of more extended UVOT photometric 
monitoring campaigns of the SFXTs were reported previously by \citet{romano11} and \citet{romano15}, for example. These authors found similar 
results, with no significant evidence of changes in the magnitudes of the supergiant stars in the SFXTs even close to the 
epoch of the different outbursts caught by \swift.\ 
Strictly simultaneous and fast (few seconds) optical, IR, and UV measurements during the X-ray outburst carried out with large telescopes 
could help investigate the presence of particularly massive clumps impacting against the neutron star. However, scheduling 
these observations is very challenging as the precise occurrence of SFXT outbursts cannot be predicted {\it a priori} 
and many periastron passages of \src\ are observed where no X-ray outburst is taking place \citep[see the discussion in, e.g.,][]{drave14}. 
The investigation of the spectral properties, 
rather than of only the photometric variations, of the SFXT supergiant companions in the optical, infrared, and ultraviolet domain has 
been shown to be able to provide some useful information on the characteristics of their stellar winds compared to those in 
classical systems \citep{gimenez16}. However, precise measurements of the wind properties are still scarce due to the large distances 
of the SFXTs that make observations challenging, especially in the ultraviolet domain. Furthermore, through the performed observational campaigns so far, it was not possible to identify 
 a net difference in the stellar wind properties of SFXTs and classical SgXBs, and 
thus the tentative discrimination between the proposed theoretical scenarios in the two classes of sources still relies upon the 
largely unknown values of the neutron star pulsation periods and magnetic field strength \citep[see discussions in][]{bozzo16,gimenez16}.

\section*{Acknowledgements}
We warmly thank the \xmm\ and \nustar\ teams for their efforts in scheduling the simultaneous 
observations of \src.\ This work was supported under NASA Contract No. NNG08FD60C, and made use of data 
from the {\it NuSTAR} mission, a project led by  the California Institute of Technology, managed 
by the Jet Propulsion  Laboratory, and funded by the National Aeronautics and Space Administration.  
This research has made use of the {\it NuSTAR} Data Analysis Software (NuSTARDAS) jointly 
developed by the ASI  Science Data Center (ASDC, Italy) and the California Institute of  Technology 
(USA). We also made use of observations obtained with \xmm,\ an ESA science mission with instruments and 
contributions directly funded by ESA Member States and NASA. 
This publication was motivated by a team meeting sponsored by the International Space Science 
Institute in Bern, Switzerland. EB, LO, and AM thank ISSI for their financial support during their   
stay in Bern. VB and JAT thank Brian Grefenstette and Kristin Madsen for help with planning the NuSTAR 
observations and Lorenzo Natalucci and David Smith for useful discussions. 
PR acknowledges contract ASI-INAF I/004/11/0 and  
financial contribution from the agreement ASI-INAF I/037/12/0. AM acknowledges support from the
Polish NCN grant 2013/08/A/ST9/00795. SC is grateful to the Centre National d’Etudes Spatiales (CNES) 
for the funding of MINE (Multi-wavelength INTEGRAL Network). We thank the anonymous referee for detailed and 
useful referee report that helped in improvement of the paper.

\bibliography{reference_17544.bib}{}
\bibliographystyle{aa}

\end{document}